\RequirePackage{fix-cm} 

\makeatletter
\def\cl@chapter{}
\makeatother

\documentclass[onecolumn]{svjour3}

\usepackage{cite}
\usepackage[colorlinks=true]{hyperref}
\usepackage{nameref}

\smartqed  

\usepackage[T1]{fontenc}
\usepackage[ngerman,english]{babel}
\usepackage{eurosym}
\usepackage{tcolorbox}
\tcbuselibrary{listings,breakable}

\usepackage{xcolor} 
\definecolor{deepblue}{rgb}{0,0,0.5}
\definecolor{deepred}{rgb}{0.6,0,0}
\definecolor{deepgreen}{rgb}{0,0.5,0}

\usepackage{booktabs}
\usepackage[table]{xcolor}
\usepackage[flushleft]{threeparttable}
\usepackage{caption}
\usepackage{subcaption}
\usepackage{multirow}

\usepackage{enumitem}

\usepackage{float}

\usepackage{graphicx}
\usepackage{rotating}
\usepackage{adjustbox}
\usepackage{tikz}
\usepackage{pgfplots}
\usepgfplotslibrary{statistics}
\usetikzlibrary{fadings}

\usepackage{amsmath,amssymb,amsfonts}
\usepackage{mathtools}
\usepackage{mathrsfs}
\usepackage{nicefrac}
\usepackage{pifont}

\usepackage{listings}

\usepackage[lined,boxruled,norelsize,linesnumbered]{algorithm2e}
\SetKwComment{Comment}{$\triangleright$\ }{}
\SetCommentSty{itshape}
\def\HiLi{\leavevmode\rlap{\hbox to \hsize{\color{gray!35}\leaders\hrule height .8\baselineskip depth .5ex\hfill}}}

\usepackage{csquotes}
\usepackage{ulem}
\usepackage{ifthen}
\usepackage{balance}
\usepackage{soul}
\usepackage{url,moreverb,xspace}

\usepackage[nohyperlinks, printonlyused, withpage]{acronym}

\usepackage[figure,table,lstlisting]{totalcount}

\newcommand{\changed}[1]{\textcolor{black}{#1}}
\newcommand{\code}[1]{\lstinline[basicstyle=\small\ttfamily, literate={`}{\textasciigrave}1, escapechar=@, upquote=true]{#1}}

\newcommand*\colourcheck[1]{\expandafter\newcommand\csname #1check\endcsname{\textcolor{#1}{\ding{52}}\xspace}}
\newcommand*\colourcross[1]{\expandafter\newcommand\csname #1cross\endcsname{\textcolor{#1}{\ding{56}}\xspace}}

\newcommand{\nb}[2]{%
  {\sf\fcolorbox{yellow}{yellow}{\scriptsize\textbf{#1}}%
  $\blacktriangleright$%
  {\color{blue}\fontsize{7pt}{8pt}\selectfont\textbf{#2}}}%
}

\newcommand{\Vera}[1]{\nb{Vera}{\hl{#1}}}
\newcommand{\Oliver}[1]{\nb{Oliver}{\hl{#1}}}

\makeatletter
\let\orgdescriptionlabel\descriptionlabel
\renewcommand*{\descriptionlabel}[1]{%
  \let\orglabel\label
  \let\label\@gobble
  \phantomsection
  \edef\@currentlabel{#1}%
  \let\label\orglabel
  \orgdescriptionlabel{#1}%
}
\makeatother

\journalname{EMSE}

\newcommand*{\myPercent}{\%}

\newcommand*{\ctb}{ansBenchmark}
\newcommand*{\ctbs}{CodeTransBenchmark }

\lstdefinestyle{mydefaultstyle}{
    basicstyle=\ttfamily\footnotesize,
    keywordstyle=\color{deepblue},
    emphstyle=\ttfamily\color{deepred},
    stringstyle=\color{deepgreen},
    frame=tb,
    showstringspaces=false,
    literate={`}{\textasciigrave}1, escapechar=@,
    upquote=true,
    float=tp,
    breaklines=true,
    breakatwhitespace=true
}

\lstdefinestyle{promptstyle}{
    basicstyle=\ttfamily\footnotesize,
    keywordstyle=\color{deepblue},
    emphstyle=\ttfamily\color{deepred},
    stringstyle=\color{deepgreen},
    frame=tb,
    showstringspaces=false,
    literate={`}{\textasciigrave}1,
    escapechar=@,
    upquote=true,
    float=tp,
    breaklines=true,
    breakatwhitespace=true,
    breakindent=5pt
}

\lstdefinestyle{pythonstyle}{
    language=Python,
    basicstyle=\ttfamily\footnotesize,
    morekeywords={self},
    keywordstyle=\color{deepblue},
    emphstyle=\ttfamily\color{deepred},
    stringstyle=\color{deepgreen},
    frame=tb,
    showstringspaces=false,
    literate={`}{\textasciigrave}1, escapechar=@,
    upquote=true,
    float=tp,
    breaklines=true,
    breakatwhitespace=true
}

\lstdefinestyle{pythonstylesmall}{
    language=Python,
    basicstyle=\ttfamily\footnotesize,
    morekeywords={self},
    keywordstyle=\ttfamily\color{deepblue},
    emphstyle=\color{deepred},
    stringstyle=\color{deepgreen},
    frame=tb,
    showstringspaces=false,
    literate={`}{\textasciigrave}1, escapechar=@,
    upquote=true,
    float=tp,
    breaklines=true,
    breakatwhitespace=true
}

\pgfplotsset{compat=1.18}
\begin{document}

\title{CodeTransBenchmark: Evaluating LLM-based Code Translation and Repair Across Programming Languages}

\author{Vera Kowalczuk \and Oliver Wei{\ss}l \and Severin Kacianka \and Andrea Stocco}

\authorrunning{Kowalczuk V. and Wei{\ss}l O. and Kacianka S. and Stocco A.}
\titlerunning{Large Language Model-based Code Translation between Programming Languages}

\institute{Vera Kowalczuk, O. Wei{\ss}l and A. Stocco are with the Technical University of Munich -- Boltzmannstra{\ss}e 3 Garching near Munich, Germany. A. Stocco is also with fortiss GmbH -- Guerickestra{\ss}e 25 Munich, Germany. S. Kacianka was with fortiss GmbH -- Guerickestra{\ss}e 25 Munich, Germany.}

\maketitle
\begin{abstract}
Large Language Models (LLMs) pre-trained on expansive text and code corpora have revealed promising code generation abilities and have attracted increasing attention in code translation. 
In this work, we investigate the effectiveness of LLMs in code translation and translation error repair. 
First, we present CodeTransBenchmark, a framework for evaluating LLM-based translation and repair and devise a post-processing strategy to extract code from inconsistent LLM outputs. 
Then, we discuss an empirical study evaluates eight models on three datasets and 12 language pairs, in which we categorize incorrect translations by errors to identify weaknesses of existing LLMs. 
Our work shows that while LLMs specifically trained for multi-lingual coding, like Codestral, correctly translate the majority of code, most general-purpose models struggle with the syntactic rules of the target language.
The analysis of erroneous translations reveals the substantial impact of the interrelationship between involved programming languages and training data on the effectiveness.
We show that a general post-processing approach must tolerate inconsistencies and leverage the predictability of LLM answers. 
Further, we show that iterative translation repair via automated feedback significantly improves translation accuracy.
While our combined findings highlight the potential of LLMs to automate code translation, an effective deployment of LLM-based code translation in practice would require models with larger context windows. 
\end{abstract}


\section{Introduction}\label{sec:introduction}

Code translation, or code migration, refers to converting source code from one programming language to another while preserving its functionality. This process has practical importance across many software engineering tasks. Maintaining a unified codebase in a modern language simplifies development, improves readability, and facilitates maintenance. Developers may also translate code to take advantage of newer language features, adapt to modern frameworks, or achieve performance, security, or scalability improvements.
A common example is prototyping in a dynamically typed language such as Python, followed by porting to a compiled language with a static type system for production deployment. Similarly, teams often translate code to enable cross-language integration, avoid abstraction layers, or unify components across projects. Code translation also plays a crucial role in porting software across operating systems and modernizing legacy or proprietary systems to align with evolving business and technological needs.

Legacy systems written in outdated languages pose particular challenges. They are often inefficient, insecure, and difficult to maintain~\cite{khan_legacy_2023}. As expertise in such languages declines~\cite{irrera_banks_2017}, migration to modern alternatives becomes essential but costly. The Commonwealth Bank of Australia, for instance, spent over 750 million USD and five years porting its platform from COBOL to Java~\cite{irrera_banks_2017}, highlighting the scale of manual translation efforts.

Early attempts to automate code translation relied on rule-based transpilers~\cite{java2csharp_java2csharp_2013, go_transpile_gotranspilecxgo_2024, shetty_crust_2019}, in which each rule mapped constructs between source and target languages. While effective in constrained domains, such systems are expensive to maintain, incomplete, and prone to producing unnatural or incorrect code~\cite{pan_lost_2024}.
Statistical machine translation (SMT) later introduced data-driven mappings~\cite{nguyen_lexical_2013, nguyen_contexts_2016, karaivanov_phrase-based_2014}, but its reliance on parallel datasets and limited context made it unsuitable for handling larger code structures. Similar constraints affected early neural methods such as CNNs, RNNs, and LSTMs~\cite{chen_tree--tree_2018}.

With the rise of Transformer architectures, encoder–decoder models trained on large monolingual code corpora achieved significant progress in unsupervised code translation~\cite{roziere_unsupervised_2020, roziere_dobf_2021, roziere_leveraging_2022, szafraniec_code_2023, liu_syntax_2023, zhu_multilingual_2022}. While these approaches improved translation quality, they remained costly to train and often unstable.
The advent of pre-trained large language models (LLMs) has shifted this landscape. LLMs trained on both text and code demonstrate strong generalization across programming tasks, including code completion~\cite{roziere_code_2024, chen_evaluating_2021}, summarization, test generation~\cite{wang_software_2023}, and debugging~\cite{fan_large_2023, hou_large_2023}. Leveraging such pre-trained models for code translation avoids retraining from scratch and offers a more resource-efficient alternative.

Recent studies have investigated how to adapt general-purpose LLMs for code translation. Yan et al.~\cite{yan_codetransocean_2023} demonstrated that prompt design substantially affects translation success. Pan et al.~\cite{pan_lost_2024} categorized common translation errors in GPT-4 outputs~\cite{openai_gpt-4_2024} and proposed iterative prompting for improvement. Macedo et al.~\cite{macedo_exploring_2024} analyzed how LLM output structures impact evaluation reliability but found no post-processing strategy that generalizes across models. Yang et al.~\cite{yang_exploring_2024} showed that incorporating test cases into prompts and repair cycles enhances translation outcomes, particularly for capable models.

\changed{Building on these findings, this paper evaluates whether open-source LLMs can serve as effective and practical alternatives to proprietary ones for
automated code translation. We focus on sized,
open-source models that can be deployed on local hardware under quantized
inference. This regime is practically important---it is the only option when cost, data privacy, or reproducibility constraints rule out proprietary APIs ---yet it is not covered by recent studies that concentrate on leading proprietary models or large-scale open models~\cite{pan_lost_2024,10.1145/3728940,gong-etal-2026-trace}. Open-source models offer advantages in cost efficiency,
reproducibility, and data privacy, as they can be deployed locally. However, they face challenges such as smaller context windows, reduced fine-tuning scope, and varying translation reliability, and---as we show---they frequently ignore output-format instructions, which complicates automated evaluation.}

To this end, we introduce CodeTransBenchmark, a framework for evaluating LLM-based code translation with integrated validation and adaptive post-processing. Through an extensive empirical study, we analyze open-source LLMs' ability to perform language-to-language translation and iteratively repair their own outputs based on automated feedback.
By expanding the range of programming languages studied and standardizing evaluation, our work contributes to developing reliable, transparent, and extensible tools for automating code migration in practice.
\changed{This paper makes the following contributions:}
\begin{itemize}
\item \changed{\textbf{Framework.} We present CodeTransBenchmark, a unified,
backend-agnostic framework for automated code translation, post-processing,
validation, and iterative repair with locally deployed LLMs. Throughout the
paper we use the term benchmark to denote this reusable evaluation
framework together with the curated evaluation datasets it operates on, rather
than a single dataset--metric pair as the term is sometimes used in the NLP
community.}
\item \changed{\textbf{Post-processing method.} We devise Flexible
Extraction, an adaptive, model-agnostic method to extract code from
inconsistent free-form LLM outputs, and show that it changes measured accuracy
substantially (by roughly 53\% relative on average) and generalizes across all
studied models.}
\item \changed{\textbf{Empirical analysis.} We evaluate eight open-source LLMs
across three datasets and twelve language pairs (67,071 translations),
systematically studying the sensitivity of translation outcomes to prompt
design, post-processing strategies, language-specific effects, the
distribution of translation errors, and single-round feedback-based repair
(15,182 regenerations).}
\end{itemize}

\changed{The remainder of this paper is organized as follows. \autoref{cap:approach} presents CodeTransBenchmark and its main components. \autoref{cap:empirical-study} describes the empirical study design, datasets, and evaluation procedure. \autoref{cap:results} reports the quantitative findings, followed by the qualitative analysis and discussion in the subsequent sections. \autoref{cap:conclusion-future-work} concludes the paper and outlines future work.}

\section{Approach}\label{cap:approach}

\begin{figure}[t]
    \centering
    \includegraphics[width=0.9\textwidth]{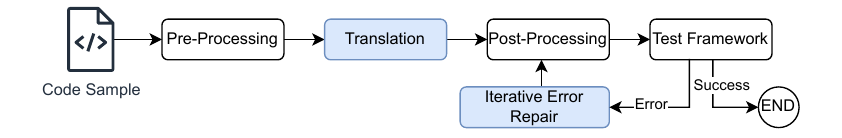}
    \caption{\changed{Overview of the \ctb{} workflow. The blue components correspond to the LLM-driven translation and repair steps.}}
    \label{fig:codetransbenchmark_system}
\end{figure}

We developed a modular and extensible LLM code translation benchmarking pipeline called \ctb. The framework was designed with a modular architecture to enable the seamless interchange of models, LLM execution frameworks, prompt strategies, and datasets.
The main components of \ctbs are shown in \autoref{fig:codetransbenchmark_system}. 
Each component represents a processing step in our pipeline: pre-processing the samples in the datasets, source code translation through LLMs, cleanup of the generated output via post-processing, and the testing framework that evaluates the correctness of the translations. 
The last step, the iterative error repair, concerns a translation error-fixing approach step.

At a high level, the components have the following purpose and connections.
As detailed in \autoref{sec:pre-processing}, the pre-processing component sanitizes the input samples in the datasets and excludes samples containing syntactic errors, incomplete code, or invalid data.
In the next step, the LLM under test is prompted to translate the sanitized code samples.
Two modules support the translation step, namely, one building the prompt to the LLM and the other abstracting from the underlying LLM and model runtime framework.
All LLM-generated code are cleaned during post-processing, and the translated code is extracted (see \autoref{sec:post-processing}).
The testing and evaluation framework compiles, executes, and tests these translations with the test cases included in the dataset, producing a report on the translation's success and feedback on errors.
We leverage these reports in our analysis and the iterative repair step, where the same LLM is tasked to improve its erroneous translations.
A central script orchestrates these components to streamline all steps involved in benchmarking LLMs on code translation.
It allows batch processing of one or multiple datasets with all pairs of programming languages.
This automation reduces manual effort and ensures consistency and replication. 
Due to the significant processing time required for inference on all translations and repairs, as well as the execution of the test framework, we included mechanisms that enable task interruption and resumption.
The remainder of this section delves into the technical specifics of all components.

\subsection{Pre-Processing}\label{sec:pre-processing}

To ensure code translation consistency, we first clean datasets of non-ASCII characters, which may affect translation quality. Concrete results about the filtered code samples are available in \autoref{tab:faulty_source_files_stats}.

\subsection{Translator}\label{sec:codetransbenchmark-translator}

\begin{figure}[t]
    \centering
    \includegraphics[width=0.9\textwidth]{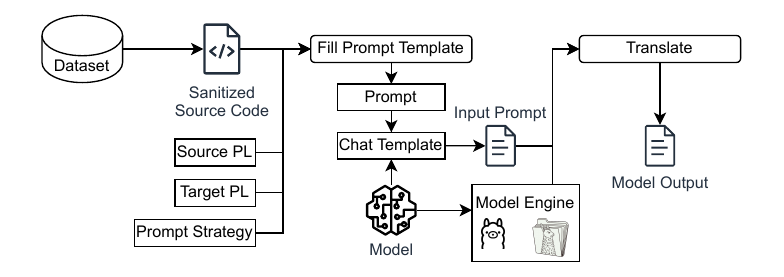}
    \caption{\changed{Translation component of \ctb. The figure shows how sanitized source code and language metadata are inserted into a prompt template, adapted to the selected model interface, and then passed to the LLM to generate a candidate translation.}}
    \label{fig:codetransbenchmark-translation}
\end{figure}

The translation component takes sanitised code samples and the specified language pair and generates a translation using a LLM. Prompt templates are filled with the source code and language metadata, then adapted to the expected input format of the target model. Before inference, the prompt is verified to fit within the model's context window (see \autoref{fig:codetransbenchmark-translation}).

Model interaction and prompt orchestration are handled through a unified abstraction layer built on LangChain~\cite{chase_langchain_2022}, which simplifies prompt management and model switching. The system supports multiple inference backends, including local llama.cpp-based engines such as Llamafile~\cite{tunney_llamafile_2024} and Ollama~\cite{ollama_ollama_2024}, as well as the HuggingFace \code{transformers} interface~\cite{wolf_huggingfaces_2020}. For our experiments, we employed llama.cpp-based models due to their efficiency and support for quantised weights, making them well-suited for local inference.

\subsection{Post-Processing}\label{sec:post-processing}

\begin{table}[t]
\centering
\scriptsize
\caption{Scoring system to distinguish code from explanations and other text.}
\label{tab:scoring_system_code_heuristic}
\begin{tabular}{@{}ll@{}}
\toprule
\textsc{Indicators for code}                      & Score            \\ \midrule
\quad Semi-colons at the end of a line   & + 1              \\
\quad Comment syntax (\code{\#}, \code{\"\"\"}, \code{//}, \code{/*})      & + 1              \\
\quad Curly braces, parentheses, or brackets                               & + 1              \\
\quad Parentheses directly following text without space in between         & + n              \\
\quad camelCase                          & + n              \\
\quad snake\_case                        & + n              \\
\quad Keywords of programming languages  & + 1 per keyword  \\
\quad Operators and comparators          & + 1 per item     \\
\quad Text indented with two spaces or tabs                                & + 1              \\ \toprule
\textsc{Indicators for explanations or comments} & Score    \\ \midrule
\quad Markdown inline code syntax        & - 2n             \\
\quad A bullet point at the beginning of the line                          & - 5n             \\
\quad An enumeration at the start of a line                                & - 5n             \\
\quad Patterns starting explanations, comments or introductions            & - 20             \\ \bottomrule
\end{tabular}
\end{table}

LLM outputs often include extraneous text alongside translations, affecting automated evaluation. Direct prompts to avoid explanations are ineffective due to instruction-tuning limitations. Instead, we propose a heuristic-based code extraction method leveraging Markdown syntax and code-specific patterns (e.g., identifiers, operators). Each line in the output is scored: code receives positive scores, natural text negative scores, and Markdown fences neutral scores (see \autoref{tab:scoring_system_code_heuristic}).
The post-processor addresses code fence imbalances (e.g., nested or missing fences) by applying the heuristic to identify code segments. For Java files, the main class is renamed to match the filename via RegEx, ensuring compiler compatibility. If no public or \code{Main} class is detected, translations are marked unsuccessful.

\begin{figure}[b]
    \centering
    \includegraphics[width=0.9\textwidth]{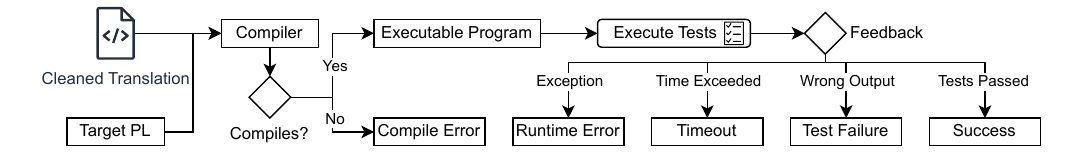}
    \caption{\changed{The test and evaluation component of CodeTransBenchmark. Each
cleaned translation is compiled (where applicable) and executed against the
dataset's test cases; outcomes are classified as compile error, runtime error
(uncaught exception or crash), timeout, test failure (wrong output), or
success.}}
    \label{fig:codetransbenchmark-testing}
\end{figure}

\subsubsection{Translation Test und Evaluation System}\label{sec:test_evaluation_pipeline}

The test and evaluation component verifies the correctness of all cleaned translations and classifies resulting errors (\autoref{fig:codetransbenchmark-testing}). It collects and reports errors from the compilation stage, the runtime environment, and functional test execution.
Each translation is compiled using the target language's compiler. If compilation is successful, the resulting executable is tested using a framework that feeds inputs to the program and compares its outputs to the expected results. A timeout mechanism ensures that non-terminating programs are flagged, and output normalization accounts for platform-specific differences.
Errors are classified into compilation, runtime, timeout, or functional errors and recorded in a structured report. Language-specific parsing is supported where necessary—for example, to filter noise in compiler messages.
\ctbs supports parallel execution of test runs and manages isolated environments to prevent interference between runs. \ctbs currently supports seven programming languages, namely C, C++, C\#, Go, Java, Python, and Rust.

\subsubsection{Translation Error Repair}\label{sec:codetransbenchmark_repair}

Other than supporting the translation task, we extended \ctbs with an iterative error correction step.
It uses the same LLM model, prompting, and inference engine abstractions as the translation component and leverages the error reports collected through the translation testing component.
As described in the previous section, the translation evaluation classifies errors as compile-time, runtime-time or test errors. 
\ctbs further collects information such as error messages by the respective compiler and runtime environment of the target language and from failing test cases.
This information is saved for each translation and retrieved for the repair process.
There, the LLM is tasked to fix its mistakes and prompted to regenerate the translation.
As mentioned in \autoref{sec:prompts_rq2}, the correct repair prompt is automatically selected depending on the error class.
With the same technology as in the translation component, the prompt is filled with the pre-processed source code, information on the programming languages involved, the erroneous translation, and the collected feedback information specified by the prompt template before inserting it in the model's chat template.
Since the repair prompts comprise the previous translations, the latter are sanitised of all non-ASCII characters in comments, using the same technique as in the pre-processing step.
\changed{It is essential to verify that the final repair prompt does not exceed the context window of the model, because it must embed the original source, the previous translation, and the collected feedback, which---for test failures in particular---can be extensive. We leverage the tokenizers of the LLMs to measure each repair prompt and skip regeneration for samples whose tokenized prompt exceeds the respective limit; error messages are never truncated, so no partial or lossy prompts are sent to the models.}



\subsubsection{Implementation Details}\label{sec:technical_setup}


We run our pipelines locally on one machine with an AMD EPYC 9354P CPU that has 32 cores and 64 logical processors, 188~GB RAM, and an 2x NVIDIA L40S with 48GB VRAM each. 

Since we run all our experiments locally and aim for a sustainable and efficient translation process, we use quantized versions of the selected models instead of full or half-precision weights. 
While all selected models fit into VRAM with our hardware, using the quantized versions reduces the required memory to load the models during inference, increases the computational efficiency of our framework, reproducibility for third parties, and speeds up the translation process.
We specifically selected the \textit{Q5\_K\_M} quantized versions of the selected models, which use a combination of five-bit and six-bit quantization resulting in 5.5 bits per weight.

\autoref{tab:hyperparameters} lists the hyperparameter settings applied throughout our study.
The temperature regulated randomness in generation, with values tending toward 0 allowing the least freedom in generation. The selection for the next token is limited to the 50 most probable tokens and the tokens with a cumulative probability above 95\myPercent. Additionally, a token needs a minimum probability of 5\myPercent{} relative to the probability of the most likely token to be considered. Since code can be somewhat repetitive, we do not penalize repeated tokens.

\begin{table}[t]
\centering
\footnotesize
\caption{Values for hyperparameters in \ctb{}.}
\label{tab:hyperparameters}
\begin{tabular}{ccccc}
\toprule
\textit{temperature} & \textit{top\_k} & \textit{top\_p} & \textit{min\_p} & \textit{repeat\_penalty} \\
 0.7         & 50     & 0.95   & 0.05   & 1               \\ \bottomrule
\end{tabular}
\end{table}

\section{Empirical Study}\label{cap:empirical-study}



\subsection{Research Questions}\label{sec:rqs}


\textbf{RQ1: How effective are open-source LLMs in translating code?} 
This question evaluates the capability of open-source LLMs to translate code. This analysis includes three sub-questions exploring critical factors influencing translation performance.

\changed{
\textbf{RQ1.1: How sensitive are translation outcomes to prompt design,
and which prompt should be selected for the main comparison?}
We systematically
compare prompting techniques including direct code-to-code translation,
explicit output-format instructions, 1-shot examples, and
code-to-text-to-code workflows. Rather than searching for a globally
``optimal'' prompt, this sub-question quantifies the stability of translation
accuracy across prompt variants and serves as a principled prompt
selection step: the best-performing variant is held fixed for all subsequent
experiments, so that the cross-model comparison in RQ1.2--RQ1.4 is not
confounded by prompt choice.
}

\textbf{RQ1.2: How well do post-processing strategies generalize across models?}
This sub-question investigates the development of universal extraction methods for source code from LLM outputs. Through cross-model analysis of output format patterns, we create and evaluate post-processing techniques that standardize code extraction for automated execution, addressing model-specific behavioral differences.

\textbf{RQ1.3: How do programming languages affect translation success?}
We analyze how language characteristics (syntax, semantics, paradigms, type systems, and verbosity) influence translation accuracy. By comparing translation outcomes across language combinations, we identify patterns in model performance that correlate with linguistic features.

\changed{\textbf{RQ1.4: How are translation errors distributed?}}
\changed{We analyse how
unsuccessful translations distribute across compilation, runtime, and
functional (test-failure) categories, using the error classification produced
by the test and evaluation component (Fig.~3). Errors are aggregated per
model, per dataset, and per language pair, which reveals where LLMs
fail during code translation and complements the aggregate success rates of
RQ1.3.}

\noindent
\textbf{RQ2: How effective are LLMs in fixing translation errors?}
We assess models' ability to correct errors iteratively using feedback from RQ1's findings. This analysis measures both the success rate of error correction and the transformation of error types during the repair process, comparing different models' error-correction capabilities shaped by their training data and fine-tuning approaches.

\subsection{Evaluation Metrics in the Study}

Several studies~\cite{roziere_unsupervised_2020, roziere_leveraging_2022, yan_codetransocean_2023, yang_exploring_2024} have compared match-based (e.g., BLEU, CodeBLEU) and execution-based metrics (e.g., Computational Accuracy, Pass@k) for code translation evaluation. As match-based metrics correlate poorly with functional correctness, in this paper we adopt execution-based metrics, which, although resource-intensive, better reflect practical code quality.

For \textbf{RQ1}, we adopt Computational Accuracy (CA) as the primary metric, defined as the proportion of translated programs that execute without errors and produce outputs identical to the source code under unit tests. CA is widely used in prior work~\cite{roziere_unsupervised_2020, roziere_dobf_2021, szafraniec_code_2023, pan_lost_2024, yang_exploring_2024, yan_codetransocean_2023, macedo_exploring_2024} and is more resource-efficient than sampling-based Pass@k. CA is reported as the ``success rate'' in this study.

For RQ1.2, we compute the rates of post-processing steps required to render translated code executable, assessing the efficacy of our post-processing framework.
For RQ1.3, we compute the error rates categorized into compile-time errors, runtime errors, and test failures (incorrect semantics), providing insights into error types and their frequencies. 
For RQ1.4, we analyse how translation errors are distributed across compilation, runtime, and functional categories to understand where LLMs fail during code translation.

\changed{For \textbf{RQ2}, our primary metric is the Debugging Success Rate
after one repair attempt (DSR@1): the proportion of samples that pass all
tests after at most one feedback-driven repair round, i.e., the initial
successes plus the previously failing translations fixed by the repair step.
DSR@$k$ generalizes this to $k$ repair rounds. The metric is
analogous to, but distinct from, pass@$k$: pass@$k$ samples $k$
independent generations and measures sampling diversity, whereas
DSR@$k$ measures corrective success given execution feedback on a
single generation. In addition, we track the incremental changes in error
rates after repair, i.e., how the distribution of errors shifts across the
compilation, runtime, and functional categories during mitigation.}

\begin{table}[t]
\caption{Comparison of the described LLMs.}
\footnotesize
\label{tab:model-selection-overview}
\resizebox{\textwidth}{!}{%
\begin{tabular}{@{}lllll@{}}
\toprule
Name                     & Parameters   & Context Size & Trained On         & Features       \\ \midrule

\textsc{Code-Related} \\

\quad \textbf{Codestral}~\cite{bartowski_bartowskicodestral-22b-v01-gguf_2024}       & 22B          & 32k            & Code               & -             \\ [0.5em]

\textsc{General-Purpose} \\

\quad \textbf{Mistral}~\cite{jobbins_theblokemistral-7b-instruct-v01-gguf_2023}         & 7B           & 8k             & SFT                & GQA, SWA       \\
\quad \textbf{Dolphin-Mistral}~\cite{jobbins_theblokedolphin-26-mistral-7b-gguf_2023} & 7B           & 16k            & + Dolphin FT       & GQA, SWA       \\
\quad \textbf{Mixtral}~\cite{jobbins_theblokemixtral-8x7b-instruct-v01-gguf_2023}        & 46.7B (12.9B) & 32k            & DPO                & SMoE, GQA, SWA \\
\quad \textbf{Dolphin-Mixtral}~\cite{jobbins_theblokedolphin-27-mixtral-8x7b-gguf_2023} & 46.7B (12.9B) & 16k            & + Dolphin FT       & SMoE, GQA, SWA \\
\quad \textbf{Dolphin-Phi-2}~\cite{jobbins_theblokedolphin-2_6-phi-2-gguf_2023}     & 2.7B         & 16k            & + Dolphin FT       & MHA            \\
\quad \textbf{Llama 3}~\cite{ollama_llama38b-instruct-q5_k_m_2024}             & 8B           & 8k             & SFT, PPO, DPO      & GQA            \\
\quad \textbf{Phi-3-mini}~\cite{quantfactory_quantfactoryphi-3-mini-4k-instruct-gguf_2024}      & 3.8B         & 4k             & Textbook, SFT, DPO & MHA            \\
\bottomrule
\end{tabular}%
}
\end{table}

\subsection{LLM Selection}

The study selected a diverse set of models optimized for code generation, instruction-following, or multi-language support. We adopted the following inclusion and exclusion criteria to ensure both technical feasibility and research relevance:

\begin{itemize}
\item \textbf{Parameter size}: Given our restricted computing capacity and 96 GB GPU VRAM constraint, models must fit within this memory limit.
\item \textbf{Context window}: A minimum 4000-token context window is required to accommodate extended prompts containing source code for translation. Smaller windows would severely limit the scope of translatable code samples.
\item \textbf{Instruction tuning}: To ensure reliable output of target language code, models must have undergone explicit instruction fine-tuning. This enables precise task interpretation despite code translation complexities.
\item \textbf{Multi-programming language training}: The code translation task demands models trained/fine-tuned on datasets containing multiple programming languages. This ensures the model can handle diverse syntax and semantic patterns across languages.
\end{itemize}

\autoref{tab:model-selection-overview} summarizes the LLM models used in this study, divided by code-related and general-purpose models. These models represent a mix of basic instruction-tuned models, code-focused fine-tuned variants (e.g., Dolphin-fine-tuned Mistral 7B v2.6, Phi-2 v2.6), and specialized coding models like Mixtral 8x7B v2.7. Their inclusion enables comparative analysis of different training approaches (general instruction tuning vs. code-specific fine-tuning) within our evaluation framework. 

Codestral~\cite{bartowski_bartowskicodestral-22b-v01-gguf_2024} was included for its code-specific training. Mistral~\cite{jobbins_theblokemistral-7b-instruct-v01-gguf_2023} and Dolphin-Mistral~\cite{jobbins_theblokedolphin-26-mistral-7b-gguf_2023} were chosen for their SFT and Dolphin fine-tuning, respectively, with extended context windows. Mixtral~\cite{jobbins_theblokemixtral-8x7b-instruct-v01-gguf_2023} and Dolphin-Mixtral~\cite{jobbins_theblokedolphin-27-mixtral-8x7b-gguf_2023} were selected for their DPO training and SMoE architecture, while Llama 3~\cite{ollama_llama38b-instruct-q5_k_m_2024} was included for its combination of SFT, PPO, and DPO. The Phi series (Phi-1, Phi-2, and Dolphin-Phi-2~\cite{jobbins_theblokedolphin-2_6-phi-2-gguf_2023}) were selected for their textbook-based training (Phi-1/2) and Dolphin fine-tuning (Dolphin-Phi-2). Phi-3-mini~\cite{quantfactory_quantfactoryphi-3-mini-4k-instruct-gguf_2024} was chosen for its SFT/DPO training and compact size. OpenCodeInterpreter was excluded due to its Python-specific focus, despite its iterative error-correction capabilities via compiler feedback~\cite{hartford_cognitivecomputationsdolphin-coder_2024}. The Dolphin variants (Mistral, Mixtral, Phi-2) were prioritized for their enhanced instruction-following and code generation through fine-tuning on high-quality datasets like Dolphin-Coder and Magicoder~\cite{hartford_dolphin_2023}. All selections aimed to evaluate performance across diverse programming languages and translation tasks, leveraging features such as GQA, SMoE, and extended context windows.

\subsection{Prompt Strategies}\label{sec:prompt_strategies}
Prompt engineering is critical to unlocking the full potential of large language models (LLMs). Research demonstrates that well-crafted system prompts enhance performance by providing clear contextual guidance~\cite{zheng_is_2023}. Crucially, assigned roles must be meaningful and task-relevant, while misleading or ambiguous information can degrade model performance~\cite{yan_codetransocean_2023}. Direct, explicit task formulations---reinforced by repeating key details like programming languages---outperform indirect instructions. Example-based prompting (1-shot/multi-shot) activates transfer learning capabilities, enabling more accurate responses.

\subsubsection{Prompts for RQ1}\label{sec:prompts_rq1}
For \textbf{RQ1}, we evaluated three strategies: the minimal prompt from Pan et al. (RP)~\cite{pan_lost_2024}, the detailed prompt from Macedo et al. (RM)~\cite{macedo_exploring_2024}, and two novel designs. First, the \textit{MD prompt} employs 1-shot prompting with explicit formatting instructions, requiring translations to be enclosed in Markdown code blocks tagged with the target language. This strategy includes a language-specific example with syntax comments to reinforce output consistency (see \autoref{lst:controlled_md_template}). Second, the \textit{VT strategy} (via text) decomposes translation into two steps: first generating a code description, then producing the final translation from that description (see \autoref{lst:via_description_translation_template_1}, \autoref{lst:via_description_translation_template_2}).

All prompts define the model's role explicitly, as system/user message separation varies across models. Each template incorporates model-specific chat formatting from technical reports, with curly-brace variables dynamically populated with translation context. The \textit{controlled prompt} in \autoref{lst:controlled_template} exemplifies this approach, combining role definition, task description, and output expectations.

\subsubsection{Prompts for RQ2}\label{sec:prompts_rq2}
To address \textbf{RQ2}, we designed error-specific repair prompts integrating: (1) original task descriptions with source code and post-processed translations; (2) error feedback from compilation, execution, or test failures. The template in \autoref{lst:compile_runtime_error_template} handles compilation/execution errors, while \autoref{lst:test_failure_io_based_template} addresses test failures by including input/output pairs from failed cases. These prompts simulate interactive sessions, concatenating historical context to guide iterative fixes. Since translations are already enclosed in Markdown code fences, repair prompts focus exclusively on error resolution without redundant formatting examples.

\changed{The repair prompt is thus selected automatically by error class:
compile-time and runtime errors are handled by the template of Listing~6,
which embeds the compiler or runtime \texttt{stderr}, whereas test failures
are handled by the template of Listing~7, which embeds the failing inputs
together with the expected and actual outputs. Operationally, the test
harness (Fig.~3) distinguishes the two as follows: a runtime error is
an uncaught exception, crash, or timeout raised while executing the compiled
(or interpreted) program, whereas a test failure is a program that runs
to completion but produces output different from the expected output. This
definition applies uniformly to statically- and dynamically-typed target
languages; the distinction is meaningful for repair because the two classes
receive different feedback (an error trace vs.\ an input/expected/actual
triple).}

\subsection{Benchmarks}\label{sec:benchmarks}
Selecting a benchmark for evaluating large language models (LLMs) in translating source code between programming languages requires careful consideration of multiple criteria to ensure comprehensive and meaningful results. The benchmark must support diverse programming language pairs across paradigms, such as functional, object-oriented, and procedural styles. A particular focus is on translating Python, Java, and Go code to Python, Java, Go, C\#, and Rust. These languages are widely used in industries like web development, enterprise software, and data science, with Python being known for its readability and dynamic typing, Java for its strict type system and object orientation, Go for its concurrency features and compiler-enforced rules, and Rust for its memory safety and ownership model. This selection allows evaluation of how well LLMs handle varying syntax complexity, type systems, and programming paradigms.




\subsubsection{Dataset Selection}\label{sec:datasets}

\changed{Code translation benchmarks vary in scope and evaluation criteria. We
surveyed the available options against two requirements: (i)~execution-based
evaluation must be possible, i.e., tasks must ship with test cases rather than
only reference translations, and (ii)~the languages of
\autoref{sec:benchmarks} must be covered as source or target languages.
Several established datasets (CodeXGLUE~\cite{lu_codexglue_2021}, 
XL-CoST~\cite{zhu_xlcost_2022}) lack sufficient test cases, limiting evaluation to
match-based metrics. Others (CodeTransOcean~\cite{yan_codetransocean_2023}, TransCoder~\cite{roziere_unsupervised_2020},
G-Trans-Eval~\cite{jiao_evaluation_2023}, HumanEval-X~\cite{zheng_codegeex_2023}, EvalPlus~\cite{liu_is_2023},
do not cover the full set of languages we study. MBXP~\cite{athiwaratkun_multi-lingual_2023}
and MultiPL-E~\cite{cassano_multipl-e_2022} rely on rule-based translations of HumanEval and MBPP
but lack reliable test coverage for validation.}

\changed{Four candidate datasets satisfy both requirements: CodeScope~\cite{yan_codescope_2023},
xCodeEval~\cite{khan_xcodeeval_2023}, CodeNet~\cite{puri_codenet_2021}, and AVATAR~\cite{ahmad_avatar_2023}. From
these candidates, we selected CodeNet and AVATAR for this study. Both use
input/output-based tests, which permits flexible selection of the target
language; AVATAR additionally provides high test coverage with multiple test
cases per snippet and parallel solutions that allow direct comparison between
Java-to-Python and Python-to-Java translation. Both derive from programming
contest problems with diverse complexities, avoid dependency-heavy
environments, minimize overlap with common training corpora, and strictly
separate training and test data. Established baselines~\cite{pan_lost_2024,
yang2024unitrans} further validate them. CodeScope and xCodeEval also met our
criteria but were not used in this study; we consider them valuable for
extending the evaluation in future work and did not include them here to keep
the experimental campaign tractable.}

\changed{In addition to these two existing datasets, we constructed
BitHacks, a third dataset of 14 translation tasks inspired by Bit
Twiddling Hacks~\cite{anderson_bit_2011}, to probe language-specific
features (type inference, bit-level operations, and API mapping) that
contest-style problems under-represent. The canonical implementations of these
algorithms are written in C; we therefore manually completed the tasks in the
studied languages and verified each against the C reference behavior and
standard-library solutions, which provides reliable ground truth for
feature-level semantics. A script automates test-case generation from
input--output pairs (Listing~8 shows the task of computing an integer's
absolute value). Because the tasks are hand-written and recent, BitHacks also
serves as a (small) probe with reduced risk of overlap with model training
corpora (see \autoref{sec:threats-to-validity}).}

\changed{In summary, the study uses three datasets: CodeNet, AVATAR, and
BitHacks. retained after sanitization. \autoref{tab:faulty_source_files_stats} reports, per dataset and source language, the samples
retained after sanitization. We note that all
tasks are at function/program-snippet granularity; class-level and multi-file
translation, which recent work shows to be considerably
harder~\cite{10.1145/3728940}, is outside the scope of this study.}

\begin{table}[t]
\footnotesize
\caption{Overview of source files that are valid or excluded from the study and the reason for exclusion.}
\label{tab:faulty_source_files_stats}
\resizebox{\textwidth}{!}{%
\begin{tabular}{@{}llllcccc@{}}
\toprule
\multicolumn{1}{c}{\multirow{2}{*}{Dataset}} & \multicolumn{1}{c}{\multirow{2}{*}{Source PL}} & \multicolumn{1}{c}{\multirow{2}{*}{Total}} & \multicolumn{1}{c}{\multirow{2}{*}{Correct}} & \multicolumn{4}{c}{Errors}                            \\ \cmidrule(l){5-8} 
\multicolumn{1}{c}{}                         & \multicolumn{1}{c}{}                           & \multicolumn{1}{c}{}                       & \multicolumn{1}{c}{}                         & Unicode & Compile & Runtime & Test Failed \\ \midrule
\multirow{3}{*}{CodeNet}                     & Go           & 200      & \textbf{195}            & 3       & 2       & 0       & 0            \\
           & Java         & 200      & \textbf{196}            & 3       & 1       & 0       & 0            \\
           & Python       & 200      & \textbf{199}            & 0       & 0       & 1       & 0            \\
\multirow{2}{*}{AVATAR}                      & Java         & 249      & \textbf{222}            & 0       & 3       & 10      & 14           \\
           & Python       & 250      & \textbf{223}            & 0       & 0       & 10      & 17           \\ \bottomrule
\end{tabular}}
\end{table}

\subsection{Procedure}\label{sec:rq_procedure}

First, we checked and sanitized all code samples across all datasets. 
For CodeNet samples, comments and docstrings are extracted using the \code{comment-parser} library~\cite{aviles_comment-parser_2022}, enabling targeted removal of non-ASCII characters while preserving code syntax. While 25 Go, 20 Java, and nine Python samples required sanitization, AVATAR files already complied with ASCII standards. Notably, six CodeNet samples retained Unicode errors post-sanitization, leading to the exclusion of 64 erroneous files. Compilation and runtime failures (27 total, primarily in AVATAR due to Python 3.9 compatibility issues) further reduced the dataset (see \autoref{tab:faulty_source_files_stats}).

Concerning RQ1, we evaluated the code translation performance of all eight selected models (Codestral, Mistral, Dolphin-Mistral, Mixtral, Dolphin-Mixtral, Dolphin-Phi-2, Phi-3-mini, and Llama 3) through the \ctbs framework on three datasets, namely the two selected datasets CodeNet and AVATAR from existing benchmarks, and our BitHacks dataset. 
For each of the datasets, we evaluated all selected programming language pairs as specified in \autoref{sec:benchmarks}, whose source language is in the dataset (twelve pairs for CodeNet, eight pairs for AVATAR, and four pairs for the BitHacks).
\changed{Across the study, these evaluated directions cover Go, Java, and Python as source languages and C\#, Go, Java, Python, and Rust as target languages, depending on dataset availability.}

To address RQ1.1, we evaluated the four translation prompts defined in \autoref{sec:prompts_rq1}, namely the two reference prompts RP and RM and our own variations MD and VT, with \ctbs on the CodeNet and AVATAR datasets.


\changed{RQ1.1 is a preliminary prompt-sensitivity study whose purpose is to select the prompt used in all subsequent experiments. We therefore evaluated a representative subset of four models spanning the study's size and training-regime spectrum (Dolphin-Mistral, Dolphin-Phi-2, Mistral, and Mixtral) rather than all eight. The evaluation was incremental: we first ran the minimal reference prompt RP on Mistral and Dolphin-Mistral to establish a baseline, then evaluated all four subset models on RM and MD. Since the prompt-chaining approach VT showed clearly worse translation success and runtime already on the two Mistral
models, we did not extend it to the remaining models. The empty cells of
\autoref{tab:prompt_templates_success} correspond exactly to these choices. The best strategy (MD) was then used for all subsequent evaluations across all eight models and all datasets, so all cross-model findings (RQ1.2--RQ1.4, RQ2) rest on the full model set under a single fixed prompt.}

To assess the post-processing generalisation in RQ1.2, we collected reports on the steps of our primary post-processing approach that are applied to the output and contributed to the final extracted translation.
To validate \ctbs heuristic's effectiveness, we compare it against three simpler approaches:

\begin{itemize}
    \item Baseline: Assumes all code is enclosed in Markdown fences, removing only fences.
    \item Naive: Removes natural text using predefined prefixes and fences.
    \item Regex-only: Extracts code within target-language fences, assuming strict format adherence.
    \item These comparisons quantify the heuristic's ability to handle diverse output patterns, ensuring robust code extraction for accurate benchmarking.
\end{itemize}

For RQ1.3, we studied differences between the evaluated programming languages and how the translation effectiveness of the models is influenced.

To answer RQ2, we utilise the translation error repair process of \ctb on all the unsuccessful translations generated with the MD prompting strategy.

\changed{For RQ2 we scope the repair study to a single model
family. We selected the five Mistral-AI-based models of our
study---Mistral, Dolphin-Mistral, Mixtral, Dolphin-Mixtral, and
Codestral---so that base architecture and tokenizer are held (approximately)
constant while the training regime varies: general instruction tuning
(Mistral), Dolphin fine-tuning (Dolphin-Mistral, Dolphin-Mixtral), a sparse
mixture-of-experts variant (Mixtral), and dedicated code training (Codestral).
This family-controlled design isolates the effect of training regime on
self-repair capability without confounding it with architectural differences
across vendors. The trade-off is that our repair findings are established for
the Mistral family only; whether they transfer to other architectures
(Llama~3, the Phi series) is an open question that we flag in
\autoref{sec:threats-to-validity} and consider a natural extension, since the repair
pipeline is fully automated.}
All models only received one attempt at rectifying their mistakes based on the evaluation and test feedback collected in RQ1.
The regenerated translations are cleaned with our devised post-processing strategy and evaluated on the test framework.
In addition to the overall translation success and resulting translation error distribution, we measured how the number of successful translations as well as the categorisation of underlying error changes.




In the course of our study, eight models and four prompt strategies were evaluated for RQ1, resulting in 67,071 translations in total.
For RQ2, 15,384 translations were regenerated, 3,839 for Mistral, 3,225 for Mixtral, 3,243 for Dolphin-Mistral, 3,243 for Dolphin-Mixtral, and 1,834 for Codestral. 

\section{Results}\label{cap:results}

\subsection{RQ1: How Effective Are Smaller Open-Source LLMs in Translating Code?}\label{sec:results-rq1}

To answer RQ1, we first evaluate the impact of different prompt strategies, then compare different post-processing approaches based on the best prompt before analysing the effect of the involved programming languages and finally considering the distribution of translation errors.

\subsubsection{RQ1.1: What Is the Impact of Various Prompt Strategies on Translation Accuracy?}\label{sec:results-rq1.1}

\autoref{tab:prompt_templates_success} demonstrates the results of the four evaluated models (Dolphin-Mistral, Dolphin-Phi-2, Mistral, and Mixtral) for the four prompt strategies in terms of CA.
These are average values for the aggregation over all language pairs in the AVATAR and CodeNet datasets.
The simple RP prompt establishes a baseline to compare the other prompts against with 17.15\myPercent{} and 7.03\myPercent{} for Dolphin-Mistral and Mistral, respectively.
For both models, the RM prompt improves the CA, and the more sophisticated MD prompt gives the best success rate at 22.66\myPercent{} and 8.53\myPercent{} for the two models, corresponding to a relative increase of 32\myPercent{} and 21\myPercent{} compared to RP.
For three models, the MD prompt leads to a slight absolute rise of 1\myPercent{} except for the smaller Dolphin-Phi-2, where the RM prompt is better.
\changed{Overall, the trend remains consistent across the evaluated models, with MD performing best in nearly all cases and only the smallest model deviating from this pattern.}
Finally, the prompt chaining method of translating code via an intermediate textual description leads to a more significant error rate with only 12.66\myPercent{} and 4.81\myPercent{} success, which is only around 55\myPercent{} of the CA of the MD prompt. 
These numbers suggest that this layer of abstraction does not result in improved results but instead introduces an additional source of error.
Based on this observation, we choose not to evaluate the other models on the VT prompt.

\begin{table}[tb]
\centering
\footnotesize
\caption{\changed{Average computational accuracy (\%) of the four prompt
strategies on AVATAR and CodeNet, for the RQ1.1 model subset. The evaluation was incremental: RP was run on Dolphin-Mistral and Mistral to establish a baseline; RM and MD were run on all four subset models; VT, after clearly underperforming on the two Mistral models, was not extended further (see Section~2.6).}}
\label{tab:prompt_templates_success}
\begin{tabular}{@{}lcccc@{}}
\toprule
Prompt & D-Mistral      & D-Phi-2       & Mistral       & Mixtral        \\ \midrule
RP     & 17.15          & -             & 7.03          & -              \\
RM     & 21.67          & \textbf{6.09} & 7.56          & 22.15          \\
MD     & \textbf{22.66} & 3.79          & \textbf{8.53} & \textbf{23.04} \\
VT     & 12.66          & -             & 4.81          & -              \\ \bottomrule
\end{tabular}
\end{table}

\begin{tcolorbox}[boxrule=0pt,sharp corners,boxsep=2pt,left=2pt,right=2pt,top=2.5pt,bottom=2pt]
\begin{center}
\begin{minipage}[t]{0.99\linewidth}
\textbf{RQ1.1}: \textit{
Prompt design substantially affects translation accuracy. Among all evaluated strategies, the MD prompt yields the highest success rates for most models, improving translation accuracy by up to one-third compared to the baseline RP prompt. \changed{The overall trend is stable across the evaluated models, with only the smallest model favoring a different prompt.} The VT prompt chaining approach performs notably worse, indicating that translating via textual descriptions introduces additional errors rather than improving accuracy.
}
\end{minipage}
\end{center}
\end{tcolorbox}

\subsubsection{RQ1.2: How Well Does Post-Processing Generalise across Different Models?}\label{sec:results-rq1.2}

To assess whether the devised post-processing strategy generalises well for all models in our evaluation, we compare it with three simpler methods.
Since the MD prompt reached the best results for most models in RQ1.1, it is used in this comparison.

\begin{table}[tb]
\centering
\footnotesize
\caption{Comparison of different post-processing strategies in terms of the CA for the MD prompt (\%).}
\label{tab:comparison_postprocessing}
\resizebox{\textwidth}{!}{%
\begin{tabular}{@{}lcccccccc@{}}
\toprule
                & Codestral & D-Mistral & D-Phi-2 & D-Mixtral & Llama 3 & Mistral & Mixtral & Phi-3 \\
Post-Processing      &           &           &         &           &         &         &         &       \\ \midrule
Remove MD            & 23.68     & 21.25     & 1.93    & 24.43     & 11.25   & 4.13    & 15.61   & 8.86  \\
Remove MD + Prefixes & 53.30     & 21.25     & 2.76    & 26.02     & 17.41   & 7.52    & 21.52   & 8.86  \\
MD Assumption        & 53.57     & 21.25     & 2.98    & 25.25     & 0.00    & 0.16    & 5.40    & 0.00  \\
Flexible Extraction  & 56.29     & 22.71     & 3.77    & 27.53     & 17.91   & 8.51    & 23.14   & 9.75  \\ \bottomrule
\end{tabular}%
}
\end{table}

\autoref{tab:comparison_postprocessing} displays the variance in CA for all eight models in our study based on all possible translations of all three datasets cleaned with the respective post-processing approaches.
The table shows that the simple removal of Markdown (MD) code formatting results in a relatively low success rate across all models.
While for two models, Dolphin-Mistral and Phi-3, there is no difference between only removing Markdown code fences and additionally using known prefixes of natural language, all other models have comments or explanations in their outputs to some extent.
This effect is the strongest for Codestral since more than half of its successful translations contain natural language text.
So, it has a CA below 25\myPercent.
This outcome suggests that only removing the code fences alone is insufficient to accurately extract code snippets from LLM outputs.
In contrast, the addition of prefix detection and removal significantly improved results for six models, with an average absolute increase of 5.94\myPercent. 
Relative to the numbers of the baseline, this makes up 42.7\myPercent.
This indicates that identifying and removing natural text based on a known list of line beginnings can be valuable in improving the automated evaluation of code translation quality.

The MD Assumption strategy assumes a specific format of the output, always prepends the opening code fence from the end of the prompt to the generation and applies a RegEx to extract code inside the resulting Markdown code blocks of the target programming language. 
The resulting CA values show that making the wrong assumptions about the structure of the output text based on the prompt can be a problem when comparing various LLMs, especially those with superficial instruction tuning.
While some models follow the assumed output format of this approach, others generate outputs that do not contain code directly followed by a closing Markdown code fence but rather make an opening remark about the code at the beginning of the generated output.
Hence, we see a reduction of the samples evaluated as successful for five of the models.
While this is only an absolute decrease of 0.75\myPercent{} compared to the combination of Markdown removal and prefix usage for Dolphin-Mixtral, this post-processing method fails to correctly extract the code from outputs of the four models that are not trained or fine-tuned for coding.
For Llama 3 and Phi-3, the resulting CA is reduced to nothing and for Mistral, only a few samples can be extracted correctly and pass their tests.
So, this approach makes assumptions that do not generalise for all benchmarked models.

Notably, Flexible Extraction outperforms the other strategies across all models, indicating that its ability to adaptively extract code from LLM outputs effectively improves the measurement of translation accuracy.
This approach applies our sophisticated multi-step strategy, which includes a scoring system to distinguish between code and additional text, automated fixing of missing Markdown code fences, and code block extraction.
The results show that the average CA improves by an absolute margin of 7.31\myPercent{} and a relative margin of approximately 53\myPercent, demonstrating a significant increase over the baseline performance.
So, the best model, Codestral, achieves an average CA of 56.29\myPercent{}.
This highlights the importance of adaptability and flexibility in post-processing and code extraction techniques.

The reasons behind these improved results lie in the steps taken during the flexible multi-step post-processing approach.

\begin{tcolorbox}[boxrule=0pt,sharp corners,boxsep=2pt,left=2pt,right=2pt,top=2.5pt,bottom=2pt]
\begin{center}
\begin{minipage}[t]{0.99\linewidth}
\textbf{RQ1.2}: \textit{
Post-processing strongly influences measured translation accuracy, and its effectiveness varies across models. Simple Markdown removal is insufficient, while adding prefix detection substantially improves results by filtering natural language text. The MD Assumption approach fails to generalize because it relies on rigid output structure assumptions. In contrast, the Flexible Extraction method consistently achieves the best performance, increasing accuracy by over 50\myPercent{} on average. This demonstrates that adaptive, model-agnostic post-processing is essential for reliable code extraction and fair evaluation across diverse LLMs.
}
\end{minipage}
\end{center}
\end{tcolorbox}

\subsubsection{RQ1.3: To What Extent Do Different Programming Languages Affect the Translation Success?}\label{sec:results-rq1.3}
\begin{figure}[t]
\centering
\includegraphics[width=0.8\textwidth]{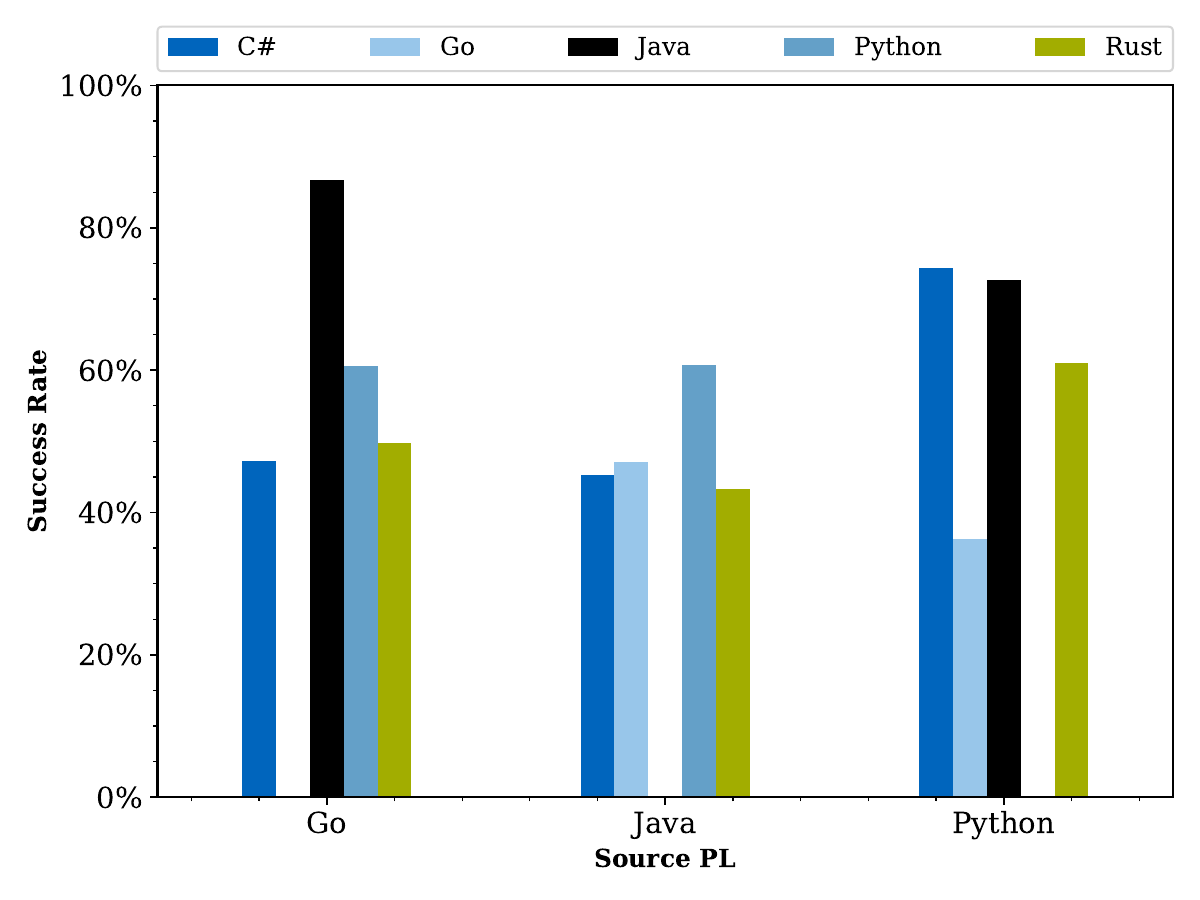}
\caption{Codestral's translation success by source and target programming language}
\label{fig:codestral_pl_combinations}
\end{figure}

To answer RQ1.3, we analyse the translation capabilities of eight LLMs across three datasets and examine how different source–target programming language combinations influence performance. The combined results are shown in \autoref{tab:iteration_1_stats_percent_total}. For each dataset, the table reports the percentage of correctly translated samples (CA) per model, along with total averages.

The results reveal substantial variation in translation success among models. The lowest-performing model, Dolphin-Phi-2, reaches only 3.77\myPercent{} CA with the MD prompt and 6.13\myPercent{} with the RM prompt, while the top-performing model, Codestral, achieves 56.29\myPercent{}, outperforming the second-best model by nearly 29 percentage points. Dolphin-Mixtral is the only other model surpassing 25\myPercent{} CA (27.53\myPercent{}). Despite its smaller size, Dolphin-Mistral performs comparably to the much larger Mixtral, suggesting that size alone does not determine translation quality.

Across datasets, all models perform worst on AVATAR, which features the highest test coverage per sample. CodeNet yields the highest success rates due to fewer tests per task, while BitHacks results lie in between. Codestral's CA varies across datasets, from 49.66\myPercent{} (AVATAR) to 61.31\myPercent{} (CodeNet). The overall ranking of average CA is: Codestral, Dolphin-Mixtral, Mixtral, Dolphin-Mistral, Llama 3, Phi-3, Mistral, and Dolphin-Phi-2.

\autoref{fig:codestral_pl_combinations} illustrates Codestral's translation success by language pair. The results show clear variation depending on both source and target languages. Translations to Java achieve the highest accuracy (around 77\myPercent{}), while those to Go are the most difficult (around 42\myPercent{}). Translating from Python tends to yield strong results, particularly for Python-to-C\# (over 74\myPercent{}).

Other models exhibit similar relative patterns but with lower overall accuracy. Dolphin-Mixtral, Mixtral, and Dolphin-Mistral follow Codestral's distribution, with Dolphin-Mixtral slightly outperforming the other two across most pairs. However, all models struggle when generating Go code, suggesting difficulties with Go's strict syntax and language rules. 
\changed{The appendix reports the corresponding per-language-pair
matrices for all eight models on all datasets. They confirm that the pattern of \autoref{fig:codestral_pl_combinations} is not specific to Codestral: Dolphin-Mixtral, Mixtral, and Dolphin-Mistral follow the same source--target distribution at lower absolute levels, and the weakest models (Llama~3, Phi-3, Dolphin-Phi-2) collapse on nearly all Go and Rust targets while retaining their best results on Java and Python targets.}

In contrast, models such as Llama 3, Phi-3, and Dolphin-Phi-2 perform poorly on nearly all Go and Rust translations. They achieve their best results when generating Java or Python code, and particularly when translating Python to C\#. These results indicate that Java and Python are generally easier target languages, while Go and Rust remain the most challenging for current LLMs.

\begin{tcolorbox}[boxrule=0pt,sharp corners,boxsep=2pt,left=2pt,right=2pt,top=2.5pt,bottom=2pt]
\begin{center}
\begin{minipage}[t]{0.99\linewidth}
\textbf{RQ1.3}: \textit{
Translation success varies considerably across models and programming languages. Codestral achieves the highest accuracy overall, while Go and Rust are the most difficult target languages due to their stricter syntax and type systems. In contrast, Java and Python are generally easier to translate from and into, likely reflecting their simpler syntax and greater representation in model training data. These results indicate that both model proficiency and language characteristics substantially influence translation performance.
}
\end{minipage}
\end{center}
\end{tcolorbox}

\begin{table}[t]
\caption{Execution metrics for all datasets, prompt templates, and models
(\%), aggregated over all language pairs. Arrows denote metric direction:
success ($\uparrow$) is higher-is-better; compile, runtime, and incorrect
($\downarrow$) are lower-is-better. Per model, the best success rate across
prompts is in bold.}
\label{tab:iteration_1_stats_percent_total}
\setlength{\tabcolsep}{2.5pt}
\renewcommand{\arraystretch}{1}
\footnotesize
\resizebox{\textwidth}{!}{%
\begin{tabular}{@{}lcccccccccccccc@{}}
\toprule
& Codestral & \multicolumn{3}{c}{D-Mistral} & \multicolumn{2}{c}{D-Phi-2} & D-Mixtral & Llama 3
& \multicolumn{3}{c}{Mistral} & \multicolumn{2}{c}{Mixtral} & Phi-3 \\
\cmidrule(l){2-2}\cmidrule(l){3-5}\cmidrule(l){6-7}\cmidrule(l){8-8}\cmidrule(l){9-9}\cmidrule(l){10-12}\cmidrule(l){13-14}\cmidrule(l){15-15}
& MD & RM & MD & VT & RM & MD & MD & MD & RM & MD & VT & RM & MD & MD \\
\midrule
\textbf{All} & & & & & & & & & & & & & & \\
\quad success\,$\uparrow$   & \textbf{56.29} & 21.66 & \textbf{22.71} & 12.56 & \textbf{6.13} & 3.77 & \textbf{27.53} & \textbf{17.91} & 7.65 & \textbf{8.51} & 4.86 & 22.31 & \textbf{23.14} & \textbf{9.75} \\
\qquad compile\,$\downarrow$ & 13.89 & 47.88 & 46.07 & 41.85 & 74.23 & 76.29 & 38.73 & 57.66 & 68.97 & 69.11 & 57.84 & 54.05 & 48.00 & 61.89 \\
\qquad runtime\,$\downarrow$ & 20.23 & 16.37 & 16.13 & 20.07 & 10.63 & 7.91 & 20.59 & 15.85 & 12.65 & 13.27 & 18.54 & 11.13 & 14.66 & 17.04 \\
\qquad incorrect\,$\downarrow$ & 9.58 & 14.08 & 15.09 & 25.52 & 9.01 & 12.04 & 13.16 & 8.57 & 10.72 & 9.10 & 18.76 & 12.51 & 14.20 & 11.32 \\
\textbf{AVATAR} & & & & & & & & & & & & & & \\
\quad success\,$\uparrow$   & 49.66 & 15.79 & 16.63 & 8.43 & 2.58 & 1.46 & 20.79 & 11.69 & 4.16 & 5.56 & 2.08 & 15.56 & 16.85 & 5.84 \\
\qquad compile\,$\downarrow$ & 17.25 & 53.37 & 51.18 & 48.26 & 78.09 & 83.71 & 42.53 & 62.53 & 75.00 & 73.31 & 63.60 & 59.66 & 53.15 & 66.29 \\
\qquad runtime\,$\downarrow$ & 19.61 & 15.51 & 15.51 & 19.66 & 10.06 & 6.46 & 21.52 & 15.39 & 11.97 & 13.48 & 18.03 & 10.96 & 14.44 & 17.30 \\
\qquad incorrect\,$\downarrow$ & 13.48 & 15.34 & 16.69 & 23.65 & 9.27 & 8.37 & 15.17 & 10.39 & 8.88 & 7.64 & 16.29 & 13.82 & 15.56 & 10.56 \\
\textbf{BitHacks} & & & & & & & & & & & & & & \\
\quad success\,$\uparrow$   & 55.36 & 21.43 & 26.79 & 5.36 & 8.93 & 1.79 & 28.57 & 25.00 & 14.29 & 7.14 & 8.93 & 33.93 & 30.36 & 8.93 \\
\qquad compile\,$\downarrow$ & 17.86 & 55.36 & 55.36 & 51.79 & 82.14 & 85.71 & 48.21 & 57.14 & 69.64 & 78.57 & 60.71 & 41.07 & 39.29 & 73.21 \\
\qquad runtime\,$\downarrow$ & 16.07 & 8.93 & 7.14 & 8.93 & 3.57 & 7.14 & 7.14 & 5.36 & 5.36 & 5.36 & 5.36 & 3.57 & 3.57 & 5.36 \\
\qquad incorrect\,$\downarrow$ & 10.71 & 14.29 & 10.71 & 33.93 & 5.36 & 5.36 & 16.07 & 12.50 & 10.71 & 8.93 & 25.00 & 21.43 & 26.79 & 12.50 \\
\textbf{CodeNet} & & & & & & & & & & & & & & \\
\quad success\,$\uparrow$   & 61.31 & 26.10 & 27.20 & 15.85 & 8.73 & 5.55 & 32.58 & 22.45 & 10.13 & 10.76 & 6.86 & 27.12 & 27.71 & 12.71 \\
\qquad compile\,$\downarrow$ & 11.27 & 43.56 & 41.99 & 36.78 & 71.13 & 70.47 & 35.64 & 54.00 & 64.41 & 65.72 & 53.43 & 50.13 & 44.32 & 58.31 \\
\qquad runtime\,$\downarrow$ & 20.81 & 17.20 & 16.82 & 20.64 & 11.23 & 9.03 & 20.21 & 16.45 & 13.35 & 13.31 & 19.24 & 11.44 & 15.08 & 17.12 \\
\qquad incorrect\,$\downarrow$ & 6.61 & 13.14 & 13.98 & 26.74 & 8.90 & 14.96 & 11.57 & 7.10 & 12.12 & 10.21 & 20.47 & 11.31 & 12.88 & 11.86 \\
\bottomrule
\end{tabular}}
\end{table}
\subsubsection{RQ1.4: How Are Translation Errors Distributed?}\label{sec:results-rq1.4}
\autoref{tab:iteration_1_stats_percent_total} shows how translation errors are distributed across compilation, runtime, and functional categories. Functional errors occur when translated programs fail tests due to changed behaviour. As described in \autoref{sec:results-rq1.3}, Codestral performs best overall, producing syntactically valid code and achieving the lowest rate of compilation errors (13.89\myPercent{}).

For most other models, compilation errors dominate. As translation success decreases, compile errors rise accordingly (for both MD and RM prompts), indicating weaker command of target language syntax. Mixtral and Dolphin-Mistral are the only exceptions, showing similar success but reversed compile error rates. Across all models, compile errors range from under 14\myPercent{} to about 75\myPercent{}, while runtime and functional errors remain stable (roughly 11–20\myPercent{} and 9–15\myPercent{}). Thus, overall success mainly depends on avoiding compile errors. For instance, Dolphin-Mixtral fails compilation nearly three times as often as Codestral.

Comparing datasets, AVATAR's lower success rate relative to CodeNet stems not only from its higher test coverage but also from more frequent compilation failures. All models show significantly more compile errors on AVATAR, which tests do not influence, and higher rates of functional errors except for the weakest models that already fail during compilation. Hence, AVATAR's lower success results from both more compilation errors and stricter test detection of semantic issues.

The prompt-chaining approach (VT) reduces compilation errors but increases logical and behavioural mismatches.

Error patterns are also language-dependent. Codestral shows high syntactic accuracy but more semantic issues: for Java, nearly all code compiles ($<4$\myPercent{} errors) yet fails at runtime (6.8\myPercent{}) or in tests (12.5\myPercent{}). Python's dynamic typing yields almost no compile errors ($<0.2$\myPercent{}) but many runtime ones (31.6\myPercent{}), often due to input parsing or API misuse. Go's strict compiler leads to 34.2\myPercent{} compile errors, Rust has a balanced split (19\myPercent{} compile, 22\myPercent{} runtime), and C\# shows mainly runtime issues (28.8\myPercent{}).

Dolphin-Mixtral produces many more compile-time errors, especially in Go (72\myPercent{}) and Rust (53\myPercent{}), while Java errors are balanced ($\sim15$\myPercent{} per category). In C\#, errors are evenly spread but more severe for Go and Java sources; for Python, compile errors are rare but runtime ones frequent.

\begin{tcolorbox}[boxrule=0pt,sharp corners,boxsep=2pt,left=2pt,right=2pt,top=2.5pt,bottom=2pt]
\begin{center}
\begin{minipage}[t]{0.99\linewidth}
\textbf{RQ1.4}: \textit{
Most translation errors occur during compilation, with runtime and functional errors remaining comparatively stable across models. Stronger models such as Codestral make fewer syntactic mistakes but more semantic ones, while weaker models fail primarily due to syntax violations. The overall success rate is therefore largely determined by how well a model can produce compilable code. Error patterns also depend on the target language: dynamically typed languages like Python show more runtime issues, whereas stricter languages like Go and Rust result in more compile-time errors.
}
\end{minipage}
\end{center}
\end{tcolorbox}


\subsection{RQ2: How Effective Are the Translation Error-Fixing Skills of LLMs?}\label{sec:results-rq2}

To address RQ2, we evaluate five of the LLMs and their capabilities to repair the mistakes they made in their translations in RQ1. 
To guide this process, they receive automated feedback based on the error messages collected during the evaluation and testing stage of \ctb.
We investigate not only the final success rate (\changed{Debugging Success Rate at 1, DSR@1}) but also the rate of improvements and analyse this in terms of the resulting and shifting distribution of errors.

\begin{table}[t]
\centering
\caption{RQ2: computational-accuracy metrics (\%) before and after one
repair round, given as \emph{before}$\rightarrow$\emph{after}. Arrows denote
metric direction: success ($\uparrow$) is higher-is-better, while compile,
runtime, and incorrect ($\downarrow$) are lower-is-better. Best success rate
per dataset in bold.}
\label{tab:stats_rq2}
\resizebox{\textwidth}{!}{%
\begin{tabular}{llcccc}
\toprule
Dataset & Model & success\,$\uparrow$ & compile\,$\downarrow$ & runtime\,$\downarrow$ & incorrect\,$\downarrow$ \\
\midrule
\multirow{5}{*}{All}
 & Codestral & \textbf{56.29$\rightarrow$70.33} & 13.89$\rightarrow$8.72  & 20.23$\rightarrow$12.94 & 9.58$\rightarrow$8.01 \\
 & D-Mistral & 22.71$\rightarrow$26.12          & 46.07$\rightarrow$51.22 & 16.13$\rightarrow$12.18 & 15.09$\rightarrow$10.49 \\
 & D-Mixtral & 27.53$\rightarrow$32.70          & 38.73$\rightarrow$33.67 & 20.59$\rightarrow$20.50 & 13.16$\rightarrow$13.13 \\
 & Mistral   & 8.51$\rightarrow$11.65           & 69.11$\rightarrow$65.25 & 13.27$\rightarrow$13.20 & 9.10$\rightarrow$9.89 \\
 & Mixtral   & 23.14$\rightarrow$33.20          & 48.00$\rightarrow$40.71 & 14.66$\rightarrow$11.99 & 14.20$\rightarrow$14.11 \\
\midrule
\multirow{5}{*}{AVATAR}
 & Codestral & \textbf{49.66$\rightarrow$66.40} & 17.25$\rightarrow$10.11 & 19.61$\rightarrow$13.26 & 13.48$\rightarrow$10.22 \\
 & D-Mistral & 16.63$\rightarrow$17.53          & 51.18$\rightarrow$70.79 & 15.51$\rightarrow$6.01  & 16.69$\rightarrow$5.67 \\
 & D-Mixtral & 20.79$\rightarrow$24.94          & 42.53$\rightarrow$38.54 & 21.52$\rightarrow$20.79 & 15.17$\rightarrow$15.73 \\
 & Mistral   & 5.56$\rightarrow$8.20            & 73.31$\rightarrow$69.21 & 13.48$\rightarrow$13.65 & 7.64$\rightarrow$8.93 \\
 & Mixtral   & 16.85$\rightarrow$24.33          & 53.15$\rightarrow$47.19 & 14.44$\rightarrow$13.54 & 15.56$\rightarrow$14.94 \\
\midrule
\multirow{5}{*}{BitHacks}
 & Codestral & \textbf{55.36$\rightarrow$64.29} & 17.86$\rightarrow$10.71 & 16.07$\rightarrow$7.14 & 10.71$\rightarrow$17.86 \\
 & D-Mistral & 26.79$\rightarrow$35.71          & 55.36$\rightarrow$35.71 & 7.14$\rightarrow$7.14  & 10.71$\rightarrow$21.43 \\
 & D-Mixtral & 28.57$\rightarrow$32.14          & 48.21$\rightarrow$44.64 & 7.14$\rightarrow$8.93  & 16.07$\rightarrow$14.29 \\
 & Mistral   & 7.14$\rightarrow$14.29           & 78.57$\rightarrow$73.21 & 5.36$\rightarrow$5.36  & 8.93$\rightarrow$7.14 \\
 & Mixtral   & 30.36$\rightarrow$46.43          & 39.29$\rightarrow$28.57 & 3.57$\rightarrow$1.79  & 26.79$\rightarrow$23.21 \\
\midrule
\multirow{5}{*}{CodeNet}
 & Codestral & \textbf{61.31$\rightarrow$73.43} & 11.27$\rightarrow$7.63  & 20.81$\rightarrow$12.84 & 6.61$\rightarrow$6.10 \\
 & D-Mistral & 27.20$\rightarrow$32.37          & 41.99$\rightarrow$36.82 & 16.82$\rightarrow$16.95 & 13.98$\rightarrow$13.86 \\
 & D-Mixtral & 32.58$\rightarrow$38.56          & 35.64$\rightarrow$29.75 & 20.21$\rightarrow$20.55 & 11.57$\rightarrow$11.14 \\
 & Mistral   & 10.76$\rightarrow$14.19          & 65.72$\rightarrow$62.08 & 13.31$\rightarrow$13.05 & 10.21$\rightarrow$10.68 \\
 & Mixtral   & 27.71$\rightarrow$39.58          & 44.32$\rightarrow$36.10 & 15.08$\rightarrow$11.06 & 12.88$\rightarrow$13.26 \\
\bottomrule
\end{tabular}%
}
\end{table}
\autoref{tab:stats_rq2} demonstrates the results of RQ2 after one round of repair, providing both the final success and error rates and the absolute improvement in per cent.

Overall, all models show an improvement in the translation success rate of over 3\myPercent. 
The LLM with the highest DSR@1 is Codestral at 70.33\myPercent, with an average improvement of over 14\myPercent{}. 
This is followed by Mixtral, which increases its success rate by 10\myPercent. 
Because this is nearly twice the repair rate of Dolphin-Mixtral, it can overtake its Dolphin-fine-tuned version by 0.5\myPercent in terms of DSR@1 (33.2\myPercent vs. 32.7\myPercent).
The models with the lowest improvement rates are Dolphin-Mistral and Mistral, with an average repair success of 3.41\myPercent{} and 3.15\myPercent{}, respectively.
While Dolphin-Mistral and Mixtral have similar success rates for the initial translation, the smaller model falls behind the larger one in DSR@1 by around 7\myPercent.

A closer look at the shifting distribution of errors can help to understand the impact of the error-fixing on the success rate.
While Codestral and the two Mixtral models can reduce all types' final number of errors, Dolphin-Mistral introduces additional compilation errors.
Mitigating an error does not necessarily lead to an increase in the success rate.
If a model manages to fix a compilation error and the translation also contains a semantic error, its error classification shifts towards a runtime or functional error.
For instance, the errors collected after the repair round reveal that many of the Go compiler errors about unused imports and variables can be mitigated by Codestral, but some programs still have logical errors.
Hence, these improvements are not directly reflected in \autoref{tab:stats_rq2}.

Part of the issue here is not only new syntactic errors introduced during the regeneration of the translation but also the token limit of the context window for the respective models.
The repair step needs an even larger context size than the initial translation; otherwise, the prompt, including the original code, translation, and error messages, exceeds this limit, and the LLM cannot generate a refined translation.

\begin{tcolorbox}[boxrule=0pt,sharp corners,boxsep=2pt,left=2pt,right=2pt,top=2.5pt,bottom=2pt]
\begin{center}
\begin{minipage}[t]{0.99\linewidth}
\textbf{RQ2}: \textit{
All evaluated models improve after one automated repair round, demonstrating that feedback-based refinement can enhance translation accuracy. Codestral achieves the largest gain, increasing its success rate by over 14\myPercent{} to 70.33\myPercent{}, while Mixtral also shows strong improvement. Smaller or less capable models such as Dolphin-Mistral and Mistral achieve only modest gains of around 3–4\myPercent{}. Error analysis reveals that repairs often shift issues from syntax to semantics rather than fully resolving them, and limited context windows can hinder effective correction. Overall, automated repair improves success but remains constrained by model capacity and context size.
}
\end{minipage}
\end{center}
\end{tcolorbox}



\subsection{Threats to Validity}\label{sec:threats-to-validity}

\subsubsection{Internal Threats} 
\changed{Two further internal threats deserve emphasis. First, all experiments
use Q5\_K\_M-quantized weights (a mixed 5/6-bit format, $\approx$5.5 bits per
weight). This choice matches the paper's target scenario of sustainable,
reproducible local inference, and Q5\_K\_M is generally regarded as close to
full precision; nevertheless, prior work shows that accuracy can vary across
quantization levels and full precision without an obvious winning
configuration, so absolute accuracies reported here may
shift under FP16 inference~\cite{Alizadeh}. We expect relative model rankings and the
qualitative error-distribution findings to be more robust than absolute CA,
but this remains an empirical question; a systematic quantization-sensitivity
study is future work. Second, generation is stochastic (temperature 0.7 with
the top-$k$/top-$p$/min-$p$ constraints of \autoref{tab:hyperparameters}), and we did not perform
repeated sampled runs per task; consequently we report point estimates without
confidence intervals or significance tests. Given the scale of the study
(67,071 translations aggregated over hundreds of samples per cell), we expect
aggregate rates to be stable, but per-pair values for small datasets (notably
BitHacks, $n$=14) should be read with caution. Adding repeated generations
with interval estimates on CA and DSR@1 is the most important methodological
extension of this work.}

\subsubsection{External Threats} 
\changed{Moreover, our tasks are at function/snippet granularity. Recent evidence
shows that LLMs perform dramatically worse on class-level translation and that
method-level benchmarks may not even differentiate models'
capabilities~\cite{10.1145/3728940}; our results should therefore be read as
an upper bound on practical translation capability for this model class.
Similarly, we evaluate functional correctness only: execution efficiency of
translated code~\cite{gong-etal-2026-trace}, code style and idiomatic
quality~\cite{zhang_instruction_2024}, and semantic equivalence beyond test
adequacy~\cite{chen_unleashing_2024} are complementary quality dimensions that our
study does not capture. Finally, the RQ2 repair findings are established for
the Mistral model family only; their transfer to other
architectures is not demonstrated here.}

\subsubsection{Evaluation and Reproducibility Threats} 
\changed{Testing-based verification cannot certify full semantic equivalence even
with adequate test suites~\cite{chen_evaluating_2021}; our CA and DSR@1 figures
are therefore relative to the datasets' test cases. Data leakage between the
public benchmark datasets and model training corpora remains possible
despite CodeNet's and AVATAR's curation; our hand-written BitHacks set (14
tasks) reduces but---given its size---cannot eliminate this risk. Constructing
larger contamination-resistant sets from recent programming tasks, as done by
recent benchmarks~\cite{yuan2024transagent}, is a stronger mitigation we plan
to adopt.}

\section{Discussion}\label{sec:discussion}


\subsection{Effective Prompting and Post-Processing for Code Translation With LLMs}\label{sec:disc-prompting-postprocessing}

Our comparison of prompting strategies shows that prompt design strongly affects translation quality.
Chaining prompts—where the model first generates a textual description before translating—introduces additional error sources and lowers accuracy.
Both steps must succeed for a correct translation, and unclear intermediate descriptions can produce invalid code.
Since code generation from text is already complex, this two-step approach compounds the difficulty.

The results also suggest that optimal prompting depends on model size.
Smaller models with limited context windows perform better with concise prompts, while longer instructions may exceed their capacity or cause confusion.
Further investigation could clarify this behaviour.

For larger models, the 1\myPercent{} CA improvement of the MD strategy over RM indicates that differences between advanced prompts are small.
However, both outperform the simple RP baseline, confirming that assigning a role and specifying output format improves outcomes.
As noted by prior work~\cite{yan_codetransocean_2023}, clear instructions and examples guide models effectively.
Our best prompt, MD, uses a one-shot Markdown example that enforces consistent output formatting and simplifies post-processing.

Since models respond differently to prompts, evaluating several options before deployment is advisable.
While prompt design can influence results~\cite{yan_codetransocean_2023, macedo_exploring_2024}, small wording changes rarely matter as long as meaning remains the same.

\noindent\textbf{Post-Processing Strategies.}
To develop a method that generalises across models, we compared simple post-processing approaches with our proposed pipeline.
Consistent with previous studies~\cite{macedo_exploring_2024}, models share output patterns but vary in format consistency.
Our Flexible Extraction approach generalises best and achieves the highest accuracy across all models.

Combining Markdown removal with prefix-based filtering improves CA for most models, showing that simple line-start rules effectively remove natural text.
In contrast, the “MD Assumption” strategy, which relies on fixed format expectations, performs poorly for models not fine-tuned for code.
Flexible Extraction adapts through a scoring system that distinguishes code from text, handling diverse outputs reliably.
A robust post-processing pipeline should tolerate format inconsistencies while leveraging predictable structure, enabling fair model comparison.

\vspace{0.3em}
\noindent\textbf{Implications.}
Effective prompting and flexible post-processing are both essential for accurate code translation.
Prompt design improves translation success, while adaptive extraction ensures fair and consistent evaluation across models.
Together, they form a unified approach to achieving reliable and comparable translation quality.

\subsection{Effectiveness of Open-Source LLMs in Translating Code}\label{sec:disc-translation-effectiveness}

LLMs show wide variation in code translation ability. Codestral performs best, achieving the highest correctness (CA) across all language pairs, suggesting strong internalisation of programming syntax and semantics. Other models translate far fewer samples and are generally unreliable.

Model size matters, but training quality and data diversity are more decisive. Although Dolphin-Mixtral is larger, Codestral—trained on code from over 80 languages—translates over twice as many samples. Code-specific fine-tuning markedly improves the Dolphin models: Dolphin-Mistral nearly matches the much larger Mixtral, while Dolphin-Phi-2 and mainly text-trained models perform worst. These findings align with external benchmarks~\cite{pan_code_2024, pan_lost_2024}, where Codestral ranks among top models, close to Claude 3 Opus~\cite{anthropic_introducing_2024}, GPT-4~\cite{openai_gpt-4_2024}, and DeepSeekCoder-33B~\cite{guo_deepseek-coder_2024}.

Translation success also depends on the programming languages involved. Go and Rust are harder targets, likely due to strict syntax and limited data, while Java and Python are easier thanks to broader representation. This asymmetry exists in both directions; effective translation requires proficiency in both source and target languages.

The AVATAR dataset reveals non-bidirectional performance, for instance: \linebreak Python$\rightarrow$Java differs from Java$\rightarrow$Python, likely reflecting differences in code structure. More parallel data is needed to confirm this.

Error analysis shows that most models fail at compile time, while stronger models like Codestral shift toward runtime and semantic errors. This indicates better syntactic understanding but incomplete semantic mastery. Common issues include input parsing and index errors, consistent with prior findings~\cite{pan_lost_2024}. AVATAR's low success rates may also stem from stricter tests and atypical code formatting.

Overall, training data quality, diversity, and fine-tuning determine translation success. Smaller open-source LLMs can approach proprietary models when trained effectively, but performance still varies by model, language pair, and task. Evaluations should therefore assess not just accuracy but also error type distributions to gauge true code understanding.
\changed{A perhaps counterintuitive observation is that translation into
Java tends to be more successful than translation into Python, despite Python's reputation as an easier language.
We hypothesize that Java's static compiler acts as a strong early filter: a
Java translation that compiles has already survived type checking, so its
residual failures concentrate in the (rarer) runtime and functional
categories. Python's dynamic typing lets more defective translations execute,
deferring their failure to runtime or to the tests---consistent with Table~7,
where Python targets show almost no compile errors ($<$0.2\%) but frequent
runtime errors (31.6\%), often due to input parsing or API misuse. Broader
representation of Java in code training corpora may contribute as well. We
present this as a hypothesis; confirming it would require a controlled
analysis on larger parallel corpora.}

\subsection{Effectiveness of Open-Source LLMs in Repairing Translation Errors}\label{sec:disc-repair-effectiveness}

\changed{All evaluated LLMs improved their translation success after one repair
round, confirming that self-correction is feasible without explicit debugging
training. We emphasize the intended role of this experiment: our repair loop
is a minimal, model-only baseline---a single round of
regeneration driven by raw compiler/test feedback, with no external test
generation, no execution alignment between source and target programs, and no
multi-agent orchestration. It quantifies how much automated feedback alone
buys for small open models, and thus establishes the floor that more
sophisticated repair pipelines should beat. Recent techniques indeed go
substantially further: UniTrans~\cite{yang2024unitrans} augments translation
and repair with generated test cases, and TransAGENT~\cite{yuan2024transagent}
coordinates four specialized agents and localizes errors through execution
alignment, reporting up to 40\% CA improvements on some pairs. Our results are
complementary: even the minimal baseline yields a 14~pp gain for Codestral
(to 70.33\% DSR@1), while the shifting error distributions and context-window
failures we observe identify precisely the bottlenecks (error localization,
context budget) that these richer pipelines address.}

The relationship between training and repair ability is non-linear. Mixtral, despite lacking code-specific fine-tuning, fixes nearly twice as many errors as its fine-tuned counterpart, Dolphin-Mixtral. Similarly, Mistral performs comparably to Dolphin-Mistral. Thus, repair effectiveness does not directly follow initial translation quality or fine-tuning scope. These results suggest that general language understanding and cross-language coding ability are sufficient for basic error correction.

Challenges remain. Regenerations can introduce new errors, and improvements may not always increase final success rates. Limited context windows often prevent inclusion of all required elements, source, translation, and error messages, restricting smaller models' repair capability. Larger windows and multi-round repair may be necessary for complex corrections.

Repair success is also language-dependent. Error message clarity, compiler behavior, and model proficiency in the target language all influence outcomes. Clear, instruction-like errors are easier to resolve than ambiguous ones. For example, Codestral handles Go compiler errors on unused imports well but struggles with logical issues requiring semantic reasoning. Parsing and highlighting error-relevant code regions could help focus model attention during repair.

In summary, LLMs can autonomously improve code translations, but effectiveness depends on model capacity, context length, and error type. Future repair strategies should combine structured error parsing, extended context handling, and iterative feedback to enhance translation correction robustness.

\changed{
\subsection{Positioning against Proprietary and large-scale Models} 
Our
study targets locally deployable open models; to let the reader
calibrate the results against the current upper bound, we contextualize them
with published numbers from studies of proprietary and large-scale
models on comparable execution-based settings. Pan et al.~\cite{pan_lost_2024}
report that GPT-4 outperforms all other studied models on C, C++, Go, Java,
and Python translations, yet even strong proprietary models show low
reliability, with success rates between 2.1\% and 47.3\% depending on the
language pair; on the associated Code Lingua leaderboard~\cite{pan_code_2024},
Codestral---our best model---ranks among the top models, close to Claude~3
Opus, GPT-4, and DeepSeekCoder-33B. More recent evaluations extend this
picture to larger model sets and harder settings: TRACE~\cite{gong-etal-2026-trace}
benchmarks 26 models including leading proprietary ones with a focus on the
execution efficiency of translated code, and ClassEval-T~\cite{10.1145/3728940}
shows that all evaluated LLMs, including DeepSeek-V3, GPT-4o, and
Claude-3.5-Sonnet, perform dramatically worse on class-level than on
method-level translation. Against this backdrop, our results should be read
as characterizing the practically relevant regime of small, open, quantized,
locally-run models on function-level tasks: Codestral's 56.29\% CA (70.33\%
DSR@1 after one repair round) falls within the reliability band reported for
much larger proprietary models on comparable function-level benchmarks, while
the remaining open models fall clearly below it. We do not claim parity with
the state of the art, and we expect the gap to widen on class-level tasks.}

\section{Related Work}\label{cap:related_work}


\subsection{Existing Techniques for Source Code Translation}\label{sec:code-translation-related-work}

Automated source code translation methods can be categorised into rule-based, statistical, and neural network-based approaches. This section summarises the primary methodologies and how LLMs relate to them.

\subsubsection{Rule-Based Transpilers}\label{rule-based-methods}

Rule-based approaches rely on converting source code into AST~\cite{mccarthy_formal_1964} and applying handcrafted rules. Tools like C2Rust~\cite{immunant_immunantc2rust_2024}, CxGo~\cite{go_transpile_gotranspilecxgo_2024}, JavaToCSharp~\cite{irwin_paulirwinjavatocsharp_2024}, and Java2CSharp~\cite{java2csharp_java2csharp_2013} operate on such principles. These tools face issues with incomplete language mappings and unidiomatic translations~\cite{pan_lost_2024}, and often require custom libraries or manual modifications. For example, C2Rust achieves 95\myPercent{} accuracy on CodeNet~\cite{pan_lost_2024} but produces unsafe or non-idiomatic Rust code~\cite{tripuramallu_towards_2024}, and does not support C++.

Improvements such as CRustS~\cite{ling_rust_2022} enhance safety and API adherence in C2Rust-generated code using additional transformation rules~\cite{emre_translating_2021}, while CRUST~\cite{shetty_crust_2019} adds support for C++ but lacks external API and module handling~\cite{tripuramallu_towards_2024}.

Transpilers also exist for specific use cases. Pyccel~\cite{bourne_pyccel_2023} translates Python to C/Fortran with human-readable output and support for scientific libraries~\cite{harris_array_2020, virtanen_scipy_2020}, while SequalsK~\cite{schultes_sequalskbidirectional_2021} enables Kotlin/Swift translation through parsing and type inference. However, all rule-based systems require deep expertise, are time-consuming to maintain, and often struggle with idiomatic output~\cite{pan_lost_2024}, especially across dynamic vs. static typing boundaries.

\subsubsection{Statistical Machine Translation (SMT)}\label{statistical-machine-translation-smt}

SMT~\cite{koehn_statistical_2009} learns statistical mappings from bilingual corpora using separate language and translation models. SMT has been adapted to code using phrase-based models like lpSMT~\cite{nguyen_lexical_2013} and tools like Phrasal~\cite{cer_phrasal_2010, green_phrasal_2014}. These approaches require curated parallel datasets~\cite{nguyen_contexts_2016} and have high lexical similarity (e.g., BLEU~81\myPercent{}) but struggle with syntax and semantics~\cite{nguyen_divide-and-conquer_2015}.

Enhancements like grammar filtering~\cite{karaivanov_phrase-based_2014} or semantic-level alignment via ASTs~\cite{nguyen_divide-and-conquer_2015} improve syntactic and semantic correctness. However, SMT still faces limitations: small context windows~\cite{phan_statistical_2020}, language-pair specificity, and dependence on data quality. SMT models cannot generalise to unseen constructs or divergent paradigms like dynamic-to-static typing~\cite{nguyen_divide-and-conquer_2015}.

Further refinements like codeSMT~\cite{nguyen_contexts_2016} include token relationships and program dependencies, boosting semantic accuracy to 80.7\myPercent. In niche use cases such as Python 2 to 3 migration, SMT tools like Moses~\cite{koehn_moses_2007} trained on 2to3~\cite{python_org_2to3_2022} yield BLEU scores above 98\myPercent{}~\cite{aggarwal_using_2015}, though still limited by n-gram context.

\subsubsection{Neural Machine Translation (NMT)}\label{neural-machine-translation-nmt}

NMT uses encoder-decoder models to map between code representations, initially via RNN, CNN, and LSTM. These models often struggle with code syntax and long-term dependencies~\cite{karpathy_visualizing_2015, dong_language_2016}.

Chen et al.~\cite{chen_tree--tree_2018} introduce a tree-to-tree model leveraging parse trees and Tree-LSTM~\cite{tai_improved_2015}, outperforming earlier approaches like lpSMT~\cite{nguyen_lexical_2013}, mppSMT~\cite{nguyen_divide-and-conquer_2015}, and Java2CSharp~\cite{java2csharp_java2csharp_2013}. Their method applies attention over tree structures, including a parent-attention mechanism. Despite improvements, limitations remain: language-pair specificity, vocabulary gaps, and performance issues with long programs or unseen code.

\subsection{LLMs and Transformer Models for Code Translation}\label{sec:llms-swe-translation}

Transformer-based models, particularly large language models (LLMs), have significantly advanced neural machine translation (NMT) for code, addressing limitations in classical NMT such as lack of parallel data and syntactic fragility. This section outlines major model families, training strategies, and recent research on LLM-based code translation.

\subsubsection{Transformer-Based Code Translation Models}\label{sec:transformer-nmt}

TransCoder~\cite{roziere_unsupervised_2020} introduced the first unsupervised transformer-based code translation model, trained on monolingual C++, Java, and Python code via cross-lingual masked language modeling, denoising autoencoding (DEA), and back-translation. Its function-level evaluation uses unit tests from GeeksForGeeks~\cite{geeksforgeeks_geeksforgeeks_2024}, introducing the Code Accuracy (CA) metric.

TransCoder-DOBF~\cite{roziere_dobf_2021} improves this by using identifier deobfuscation as a pretraining task. TransCoder-ST~\cite{roziere_leveraging_2022} filters noisy back-translations through EvoSuite-generated unit tests~\cite{fraser_evosuite_2011}, using only passing translations for self-training. TransCoder-IR~\cite{szafraniec_code_2023} adds LLVM-based intermediate representations~\cite{lattner_llvm_2004} during training, achieving up to 26\myPercent{} accuracy gains for Rust.

\subsubsection{Pretrained Multitask Models for Code}\label{sec:general-purpose-transformers}

Several pretrained transformer models support translation as part of broader code understanding:

\begin{itemize}
    \item \textbf{CodeBERT}~\cite{feng_codebert_2020}, based on BERT~\cite{devlin_bert_2019}, uses masked language modeling but lacks structural awareness.
    \item \textbf{GraphCodeBERT}~\cite{guo_graphcodebert_2021} adds data-flow edges to represent variable dependencies.
    \item \textbf{CodeT5}~\cite{wang_codet5_2021} incorporates identifier and comment semantics, and outperforms earlier models on CodeXGLUE~\cite{lu_codexglue_2021}.
    \item \textbf{PLBART}~\cite{ahmad_unified_2021} adapts BART's seq2seq architecture to source code and natural text using DEA, improving Java–C\# translation over SMT~\cite{koehn_moses_2007}.
\end{itemize}

\subsubsection{Syntax-Aware and Low-Resource Enhancements}\label{sec:syntax-aware}

SDA-Trans~\cite{liu_syntax_2023} incorporates graph attention over AST-based dependency graphs, yielding results competitive with TransCoder on Java–Python despite using a smaller dataset. Zhu et al.~\cite{zhu_multilingual_2022} propose CoST, a multilingual parallel dataset of code snippets in seven languages, and fine-tune TransCoder-DOBF on snippet-level DEA and translation. This improves BLEU and CodeBLEU scores, particularly for low-resource language pairs.

\subsubsection{Prompting and Evaluation Strategies for LLM Code Translation}\label{sec:llm-evaluation}

Studies such as CodeTransOcean highlight prompt design and hyperparameter tuning for LLM translation across 47 languages. Explicit prompts outperform role-based or chain-of-thought ones, while temperature and top-k have minimal impact. Splitting code for translation reduces accuracy, but one-shot examples provide small gains. Execution-based self-debugging strategies yield moderate improvements, as shown by Pan et al.~\cite{pan_lost_2024}.

These studies demonstrate that LLM-based translation remains sensitive to prompt formulation, output formatting, and evaluation design. While LLMs produce more human-like and context-aware code than traditional approaches, they continue to struggle with syntactic correctness, semantic preservation, and dependency resolution in real-world projects.

\subsubsection{Evaluation Benchmarks and Metrics}

Pan et al.~\cite{pan_lost_2024} compare seven LLMs (e.g., GPT-4, CodeGeeX, Llama 2) on C, C++, Go, Java, and Python translations using compilation, execution, and unit test success as metrics. GPT-4 outperforms all others but even strong models show low reliability (2.1–47.3\myPercent{} success). An iterative error-feedback prompting strategy improves accuracy by 5.5\myPercent{}, but may introduce new bugs.

Qi et al.~\cite{qi_sut_2023} introduce SUT, a benchmark for syntactic compliance, linking unit test failures to syntax errors for Java, Python, C++, and C\#. Their metrics (SUT accuracy, Syntax Element Test Scores) expose issues in operator use and variable declarations, guiding targeted fine-tuning.

\changed{With this work, we complement prior research by providing a comprehensive
evaluation of modestly sized open-source LLMs across twelve language pairs and
three benchmark datasets, under the cost-, privacy-, and
reproducibility-constrained setting of quantized local inference. In contrast
to existing studies that focus predominantly on large-scale or proprietary
models, we introduce a reusable evaluation framework,
CodeTransBenchmark, along with Flexible Extraction, a robust
post-processing strategy to extract valid outputs from inconsistent model
generations---a necessity in this regime, where models frequently ignore
output-format instructions. Our categorization of translation errors offers
insight into the influence of language-pair characteristics and model training
on translation quality, and our single-round feedback-based repair establishes
a minimal self-correction baseline (Section~4.3) against which richer
pipelines such as UniTrans and TransAGENT can be measured.}

\changed{\subsubsection{Recent Directions} 
The evaluation landscape has recently broadened
along several axes. TransAGENT~\cite{yuan2024transagent} improves LLM-based
translation with a four-agent pipeline (translation, syntax fixing, code
alignment, semantic fixing) whose key insight is error localization via
execution alignment between source and target programs, and constructs its
benchmark from recent tasks to mitigate training-data contamination.
UniTrans~\cite{yang2024unitrans} augments translation and repair with
automatically generated test cases across GPT-3.5 and LLaMA models of diverse
sizes. ClassEval-T~\cite{10.1145/3728940} lifts evaluation to class-level
translation and shows that all studied LLMs---proprietary and open---degrade
sharply relative to method level. TRACE~\cite{gong-etal-2026-trace} benchmarks the
execution efficiency of translated code across 26 models, a dimension
orthogonal to functional correctness. F2STRANS~\cite{zhang2025functiontostyle}
introduces function-to-style guidance and style-aware metrics (CCSim) over a
newly constructed benchmark with ground-truth translations. Finally,
Cheung~\cite{chen_unleashing_2024} argues that testing-based verification is
fundamentally insufficient for translation correctness and advocates formal
compositional reasoning. These works target larger or proprietary models,
richer repair pipelines, or additional quality dimensions; none addresses the
regime studied here---small open models under quantized local inference with
robust output extraction---which is precisely where evaluation-infrastructure
concerns (format non-compliance, context limits) dominate.}

\section{Conclusions and Future Work}\label{cap:conclusion-future-work}

This work explored the potential of Large Language Models (LLMs) for translating source code between programming languages, with a focus on open-source models. We developed an extensible benchmarking framework that performs translation, post-processing, evaluation, and iterative repair of code using LLMs.

Through an empirical study involving eight LLMs, three datasets, and twelve language pairs, we found that while most models have still limitations, with the exception of code-specific LLMs which can translate many self-contained examples with reasonable accuracy. Success depends heavily on the language pair, model training, prompt design, and post-processing strategies.

Our results highlight the need for adaptable output extraction pipelines and well-crafted prompts. A significant portion of translation failures stemmed from syntax errors in the target language, suggesting limited cross-language coding knowledge. We also showed that combining translation with automated feedback loops enables models to repair many errors, though challenges remain in error specificity and context size limitations.

Several promising directions emerge from this study. First, future research could devise evaluation metrics that assess both functional correctness and idiomatic quality of translated code. Further prompt engineering, including model-specific and language-aware prompts, could improve translation accuracy.
Model performance might also be enhanced by tuning generation hyper-parameters or enriching repair prompts with targeted error context. Parsing and filtering error messages and marking faulty lines can better focus model attention during repair.
More capable small models with longer context windows are needed to handle real-world code, which typically includes external libraries, multi-file dependencies, and complex language features.
To address this, agent-based translation systems may become useful, decomposing the task into modular components such as syntax analysis, translation, and repair. Refining these approaches will advance the deployment of LLMs for practical, accurate, and scalable code translation.

\changed{Another promising direction is grammar-constrained (structured) decoding:
where the inference backend supports it (e.g., GBNF grammars in
\texttt{llama.cpp} or libraries such as \texttt{outlines}), forcing the model
to emit code-only or JSON-structured outputs could reduce the post-processing
burden; comparing free-form generation with constrained decoding---and
measuring whether constraints affect translation quality itself---is a natural
follow-up to our post-processing study. Likewise, a quantization-sensitivity
study (full precision vs.\ several quantization levels)
and repeated-sampling runs with interval estimates would strengthen the
statistical footing of the reported results.}

\section{Declarations} 


\subsection{Funding}
This research was funded by the Bavarian Ministry of Economic Affairs, Regional Development and Energy. 

\subsection{Ethical Approval}  
Not applicable.

\subsection{Informed Consent}  
Not applicable.

\subsection{Author Contributions}  

\textbf{Vera Kowalczuk}: conceptualization, methodology, implementation, evaluation, writing, review, editing. \textbf{Oliver Wei{\ss}l}: implementation, evaluation, writing, review, editing.
\textbf{Severin Kacianka}: review, editing.
\textbf{Andrea Stocco}: conceptualization, methodology, review, editing.

\subsection{Data Availability Statement}  

All our results, the source code, and the simulator are accessible and can be reproduced~\cite{replication-package}. 

\subsection{Conflict of Interest}  
The authors declare no conflict of interest.

\subsection{Clinical Trial Registration}  
Clinical trial number: Not applicable.

\balance
\bibliographystyle{spmpsci}
\bibliography{literature}

\newpage
\appendix
\section{Empirical Study}

\begin{lstlisting}[caption=RP: Reference prompt of Pan et al.~\cite{pan_lost_2024}, label=lst:lit_template, style=promptstyle]
{source_pl} Code:

{source_code}

Translate the above {source_pl} code to {target_pl}.

{target_pl} Code:

\end{lstlisting}


\begin{lstlisting}[caption=RM: Reference prompt of Macedo et al.~\cite{macedo_exploring_2024}~\cite{macedo_exploring_2024}, label=lst:controlled_template, style=promptstyle]
You are a skilled software developer proficient in multiple programming languages. Your task is to re-write the input source code. Below is the input source code written in {source_pl} that you should re-write into {target_pl} programming language. You must respond with the {target_pl} output code only.

Source code:
```
{source_code}
```

### Response:
```  
\end{lstlisting}


\begin{lstlisting}[caption=MD prompt based on 1-shot learning, label=lst:controlled_md_template, style=promptstyle]
You are a skilled software developer proficient in multiple programming languages. Your task is to re-write the input source code. Below is the input source code written in {source_pl} that you should re-write into {target_pl} programming language. You must respond with the {target_pl} output code only placed in a Markdown code block.
A Markdown code block for {target_pl} looks like this:
```{target_pl_md}
{target_pl_comment} Example code
```

Source code:
```
{source_code}
```

### Response:
```{target_pl_md}
\end{lstlisting}


\begin{lstlisting}[caption=VT approach: VT Code to description prompt, label=lst:via_description_translation_template_1, style=promptstyle]
You are a skilled software developer proficient in multiple programming languages. The goal is to re-write source code in a different programming language. Describe the following source code written in {source_pl}: 

Source code:
```
{source_code}
```

### Response:
\end{lstlisting}


\begin{lstlisting}[caption=VT approach: Description to translated code prompt, label=lst:via_description_translation_template_2, style=promptstyle]
You are a skilled software developer proficient in multiple programming languages. Your task is to write a program following the given description. Below is the description of the source code that you should re-write into {target_pl} programming language. You must respond with the {target_pl} output code only. 

Description:
{description} 

### Response:
```
\end{lstlisting}










\begin{lstlisting}[caption=Error-fixing prompt for compile and runtime errors; the \{error\_class\} variable is instantiated with the detected error class, label=lst:compile_runtime_error_template, style=promptstyle]
You are a skilled software developer proficient in multiple programming languages. Your task is to re-write the input source code. Below is the input source code written in {source_pl} that you should re-write into {target_pl} programming language. You must respond with the {target_pl} output code only placed in a Markdown code block.

Source code:
```
{source_code}
```

### Response:
```{target_pl_md}
{translated_code}
```

Executing your generated code gives the following {error_class} error.

Error message:

{stderr}

Please re-generate your response and translate the above {source_pl} code to {target_pl}. You must respond with the {target_pl} output code only placed in a Markdown code block. Make sure your generated code is syntactically correct.

### Response:
```{target_pl_md}
\end{lstlisting}








\begin{lstlisting}[caption=Error-fixing prompt for test failures, label=lst:test_failure_io_based_template, style=promptstyle]
You are a skilled software developer proficient in multiple programming languages. Your task is to re-write the input source code. Below is the input source code written in {source_pl} that you should re-write into {target_pl} programming language. You must respond with the {target_pl} output code only placed in a Markdown code block.

Source code:
```
{source_code}
```

### Response:
```{target_pl_md}
{translated_code}
```

Executing your generated code gives the following output:
{generated_output}

instead of the following expected output:
{test_outputs}

Please re-generate your response and translate the above {source_pl} code to {target_pl}. You must respond with the {target_pl} output code only placed in a Markdown code block and keep the method signature from the incorrect translation. Make sure your generated code is syntactically correct. Your generated {target_pl} code should take the following input and generate the expected output:

Input:
{test_inputs}

Expected Output:
{test_outputs}

### Response:
```{target_pl_md}
\end{lstlisting}

\begin{lstlisting}[language=Python, style=pythonstylesmall, caption=Sample number three of the BitHacks dataset., label=lst:BitHacks_003]
import sys

CHAR_BIT = sys.int_info.bits_per_digit
SIZE_INT = sys.int_info.sizeof_digit


def abs_int(v: int):
    mask = v >> (SIZE_INT * (CHAR_BIT - 1))
    r = (v + mask) ^ mask
    return r


def main():
    try:
        v = int(input())
    except ValueError:
        print("Invalid input. Please enter an integer.")
        return

    result1 = abs_int(v)
    reference = abs(v)

    if result1 == reference:
        print(result1)
    else:
        print("The functions returned different results:", reference, "and", result1)


if __name__ == "__main__":
    main()
\end{lstlisting}
\begin{lstlisting}[language=Python, basicstyle=\ttfamily\footnotesize, caption=Example for the code extraction heuristic, label=lst:code_heuristic_example, style=pythonstylesmall]
-20 || Here is the Python code equivalent to the given Java code:
  0 || ```python
  1 || import sys
  0 || 
  4 || n, m = map(int, input().split())
  8 || A = [list(map(int, input().split())) for x in range(n)]
  5 || b = list(map(int, input().split()))
  0 ||
  2 || for i in range(n):
  2 ||     c = 0
  3 ||     for j in range(m):
  5 ||         c += A[i][j] * b[j]
  2 ||     print(c)
  0 || ```
-20 || Explanation:
  0 ||
 -6 || * We use the `input()` function to read input from stdin, just like how `Scanner.nextInt()` is used in Java.
 -3 || * The `map()` function is used to convert all elements in a list (or iterable) to a specified type. In this case, we use it to convert the strings returned by `input().split()` to integers.
 -8 || * We create a 2D list `A` using list comprehension and initialise `b` as a list of integers.
 -4 || * The nested loops for calculating the dot product are similar to the Java code, except we use `range(n)` instead of `i < n`, and `range(m)` instead of `j < m`.
 -5 || * We print the result using the `print()` function, which automatically adds a newline character at the end.
\end{lstlisting}

\section{Per-language Results}


\begin{table*}[t]
\centering
\caption{Detailed execution metrics for all datasets, prompt templates, and models for translations from Go to C\#}
\label{tab:go-to-csharp-execution-metrics}
\resizebox{\textwidth}{!}{%
\begin{tabular}{llrrrrrrrrrrrrrr}
\toprule
& & Codestral & \multicolumn{3}{c}{D-Mistral} & \multicolumn{2}{c}{D-Phi-2} & D-Mixtral & Llama 3 & \multicolumn{3}{c}{Mistral} & \multicolumn{2}{c}{Mixtral} & Phi-3 \\
\cmidrule(lr){3-3}\cmidrule(lr){4-6}\cmidrule(lr){7-8}\cmidrule(lr){9-9}\cmidrule(lr){10-10}\cmidrule(lr){11-13}\cmidrule(lr){14-15}\cmidrule(lr){16-16}
& & MD & RM & MD & VT & RM & MD & MD & MD & RM & MD & VT & RM & MD & MD \\
\midrule
\multicolumn{2}{l}{\textbf{All}} \\
& success   & 47.18 & 12.31 & 17.44 & 12.82 &  3.59 &  7.18 & 10.77 &  3.08 &  7.18 &  3.59 &  3.59 & 12.82 &  9.23 &  1.03 \\
& compile   &  4.62 & 33.33 & 33.33 & 22.05 & 67.69 & 67.69 & 37.44 & 76.41 & 71.28 & 67.69 & 45.64 & 62.05 & 58.97 & 56.41 \\
& runtime   & 45.13 & 33.33 & 26.15 & 28.72 & 16.41 & 14.36 & 32.31 & 16.92 & 11.79 & 14.36 & 26.67 & 13.85 & 18.46 & 13.33 \\
& incorrect &  3.08 & 21.03 & 23.08 & 36.41 & 12.31 & 10.77 & 19.49 &  3.59 &  9.74 & 14.36 & 24.10 & 11.28 & 13.33 & 29.23 \\
\addlinespace
\multicolumn{2}{l}{\textbf{CodeNet}} \\
& success   & \textbf{47.18} & 12.31 & 17.44 & 12.82 &  3.59 &  7.18 & 10.77 &  3.08 &  7.18 &  3.59 &  3.59 & 12.82 &  9.23 &  1.03 \\
& compile   &  4.62 & 33.33 & 33.33 & 22.05 & 67.69 & 67.69 & 37.44 & 76.41 & 71.28 & 67.69 & 45.64 & 62.05 & 58.97 & 56.41 \\
& runtime   & 45.13 & 33.33 & 26.15 & 28.72 & 16.41 & 14.36 & 32.31 & 16.92 & 11.79 & 14.36 & 26.67 & 13.85 & 18.46 & 13.33 \\
& incorrect &  3.08 & 21.03 & 23.08 & 36.41 & 12.31 & 10.77 & 19.49 &  3.59 &  9.74 & 14.36 & 24.10 & 11.28 & 13.33 & 29.23 \\
\bottomrule
\end{tabular}%
}
\end{table*}

\begin{table*}[t]
\centering
\caption{Detailed execution metrics for all datasets, prompt templates, and models for translations from Go to Java}
\label{tab:go-to-java-execution-metrics}
\resizebox{\textwidth}{!}{%
\begin{tabular}{llrrrrrrrrrrrrrr}
\toprule
& & Codestral & \multicolumn{3}{c}{D-Mistral} & \multicolumn{2}{c}{D-Phi-2} & D-Mixtral & Llama 3 & \multicolumn{3}{c}{Mistral} & \multicolumn{2}{c}{Mixtral} & Phi-3 \\
\cmidrule(lr){3-3}\cmidrule(lr){4-6}\cmidrule(lr){7-8}\cmidrule(lr){9-9}\cmidrule(lr){10-10}\cmidrule(lr){11-13}\cmidrule(lr){14-15}\cmidrule(lr){16-16}
& & MD & RM & MD & VT & RM & MD & MD & MD & RM & MD & VT & RM & MD & MD \\
\midrule
\multicolumn{2}{l}{\textbf{All}} \\
& success   & 86.67 & 44.62 & 55.38 & 24.10 &  9.79 & 11.79 & 57.95 & 36.41 &  7.69 & 15.38 & 15.90 & 52.82 & 47.69 & 11.79 \\
& compile   &  2.56 & 28.72 & 20.51 & 27.69 & 72.16 & 57.95 & 15.90 & 37.95 & 55.38 & 66.15 & 36.92 & 27.69 & 24.10 & 55.90 \\
& runtime   &  3.59 & 18.46 & 12.31 & 14.87 & 11.86 & 13.33 & 15.38 & 18.97 & 17.95 & 10.26 & 13.85 &  9.74 & 16.41 & 24.62 \\
& incorrect &  7.18 &  8.21 & 11.79 & 33.33 &  6.19 & 16.92 & 10.77 &  6.67 & 18.97 &  8.21 & 33.33 &  9.74 & 11.79 &  7.69 \\
\addlinespace
\multicolumn{2}{l}{\textbf{CodeNet}} \\
& success   & 86.67 & 44.62 & 55.38 & 24.10 &  9.79 & 11.79 & 57.95 & 36.41 &  7.69 & 15.38 & 15.90 & 52.82 & 47.69 & 11.79 \\
& compile   &  2.56 & 28.72 & 20.51 & 27.69 & 72.16 & 57.95 & 15.90 & 37.95 & 55.38 & 66.15 & 36.92 & 27.69 & 24.10 & 55.90 \\
& runtime   &  3.59 & 18.46 & 12.31 & 14.87 & 11.86 & 13.33 & 15.38 & 18.97 & 17.95 & 10.26 & 13.85 &  9.74 & 16.41 & 24.62 \\
& incorrect &  7.18 &  8.21 & 11.79 & 33.33 &  6.19 & 16.92 & 10.77 &  6.67 & 18.97 &  8.21 & 33.33 &  9.74 & 11.79 &  7.69 \\
\bottomrule
\end{tabular}%
}
\end{table*}

\begin{table*}[t]
\centering
\caption{Detailed execution metrics for all datasets, prompt templates, and models for translations from Go to Python}
\label{tab:go-to-python-execution-metrics}
\resizebox{\textwidth}{!}{%
\begin{tabular}{llrrrrrrrrrrrrrr}
\toprule
& & Codestral & \multicolumn{3}{c}{D-Mistral} & \multicolumn{2}{c}{D-Phi-2} & D-Mixtral & Llama 3 & \multicolumn{3}{c}{Mistral} & \multicolumn{2}{c}{Mixtral} & Phi-3 \\
\cmidrule(lr){3-3}\cmidrule(lr){4-6}\cmidrule(lr){7-8}\cmidrule(lr){9-9}\cmidrule(lr){10-10}\cmidrule(lr){11-13}\cmidrule(lr){14-15}\cmidrule(lr){16-16}
& & MD & RM & MD & VT & RM & MD & MD & MD & RM & MD & VT & RM & MD & MD \\
\midrule
\multicolumn{2}{l}{\textbf{All}} \\
& success   & 60.51 & 35.90 & 38.97 & 25.64 & 28.21 & 11.28 & 41.03 & 38.46 & 22.05 & 16.41 & 14.87 & 27.69 & 26.67 & 27.69 \\
& compile   &  0.00 &  5.13 &  9.23 &  9.74 & 28.72 & 24.10 & 12.31 &  5.64 & 24.10 & 27.69 & 12.31 & 37.44 & 30.26 &  9.74 \\
& runtime   & 35.38 & 40.00 & 35.38 & 40.51 & 26.67 & 16.92 & 37.44 & 47.18 & 36.41 & 42.56 & 38.97 & 18.97 & 26.15 & 49.23 \\
& incorrect &  4.10 & 18.97 & 16.41 & 24.10 & 16.41 & 47.69 &  9.23 &  8.72 & 17.44 & 13.33 & 33.85 & 15.90 & 16.92 & 13.33 \\
\addlinespace
\multicolumn{2}{l}{\textbf{CodeNet}} \\
& success   & 60.51 & 35.90 & 38.97 & 25.64 & 28.21 & 11.28 & 41.03 & 38.46 & 22.05 & 16.41 & 14.87 & 27.69 & 26.67 & 27.69 \\
& compile   &  0.00 &  5.13 &  9.23 &  9.74 & 28.72 & 24.10 & 12.31 &  5.64 & 24.10 & 27.69 & 12.31 & 37.44 & 30.26 &  9.74 \\
& runtime   & 35.38 & 40.00 & 35.38 & 40.51 & 26.67 & 16.92 & 37.44 & 47.18 & 36.41 & 42.56 & 38.97 & 18.97 & 26.15 & 49.23 \\
& incorrect &  4.10 & 18.97 & 16.41 & 24.10 & 16.41 & 47.69 &  9.23 &  8.72 & 17.44 & 13.33 & 33.85 & 15.90 & 16.92 & 13.33 \\
\bottomrule
\end{tabular}%
}
\end{table*}


\begin{table*}[t]
\centering
\scriptsize
\caption{Detailed execution metrics for all datasets, prompt templates, and models for translations from Go to Rust.}
\resizebox{\textwidth}{!}{%
\begin{tabular}{llrrrrrrrrrrrrrr}
\toprule
Dataset & Metric & Codestral & D-Mistral & D-Phi-2 & D-Mixtral & Llama3 & Mistral & Mixtral & Phi-3 \\
& & MD & RM & MD & VT & RM & MD & MD & MD & RM & MD & VT & RM & MD & MD\\
\midrule
All & success &49.74&14.36&12.31&15.90&0.51&0.51&15.90&8.72&0.51&1.54&1.54&6.67&8.21&0.51\\
& compile &20.51&72.82&73.33&60.00&94.36&93.33&60.51&70.26&96.41&94.36&89.23&80.00&71.28&97.95\\
& runtime &26.67&5.13&8.21&18.46&0.00&0.51&19.49&14.36&0.51&1.03&5.13&6.67&9.23&0.00\\
& incorrect &3.08&7.69&6.15&5.64&5.13&5.64&4.10&6.67&2.56&3.08&4.10&6.67&11.28&1.54\\
\midrule
CodeNet & success &49.74&14.36&12.31&15.90&0.51&0.51&15.90&8.72&0.51&1.54&1.54&6.67&8.21&0.51\\
& compile &20.51&72.82&73.33&60.00&94.36&93.33&60.51&70.26&96.41&94.36&89.23&80.00&71.28&97.95\\
& runtime &26.67&5.13&8.21&18.46&0.00&0.51&19.49&14.36&0.51&1.03&5.13&6.67&9.23&0.00\\
& incorrect &3.08&7.69&6.15&5.64&5.13&5.64&4.10&6.67&2.56&3.08&4.10&6.67&11.28&1.54\\
\bottomrule
\end{tabular}
}
\end{table*}

\begin{table*}[t]
\centering
\scriptsize
\caption{Detailed execution metrics for all datasets, prompt templates, and models for translations from Java to C\#.}
\resizebox{\textwidth}{!}{%
\begin{tabular}{llrrrrrrrrrrrrrr}
\toprule
Dataset & Metric & Codestral & D-Mistral & D-Phi-2 & D-Mixtral & Llama3 & Mistral & Mixtral & Phi-3 \\
& & MD & RM & MD & VT & RM & MD & MD & MD & RM & MD & VT & RM & MD & MD\\
\midrule
All & success &45.22&15.79&16.27&11.72&1.44&1.20&11.48&5.02&2.39&3.59&3.35&5.74&8.37&1.91\\
& compile &7.89&49.28&53.83&23.44&82.30&84.45&35.65&82.06&81.58&83.01&42.82&72.01&62.44&49.28\\
& runtime &43.30&25.36&20.57&34.93&8.61&6.70&35.89&10.77&12.20&9.09&36.84&11.24&16.27&32.78\\
& incorrect &3.59&9.57&9.33&29.90&7.66&7.66&16.99&2.15&3.83&4.31&16.99&11.00&12.92&16.03\\
\midrule
AVATAR & success &48.20&22.97&20.27&10.81&0.45&0.00&14.41&9.46&2.25&5.86&2.25&5.86&9.46&1.80\\
& compile &5.41&43.24&52.25&27.48&86.94&89.19&33.78&67.57&84.23&81.53&47.75&79.73&71.62&48.20\\
& runtime &43.24&29.73&21.17&35.59&7.21&5.41&39.64&19.82&11.26&10.36&36.04&8.11&12.16&37.39\\
& incorrect &3.15&4.05&6.31&26.13&5.41&5.41&12.16&3.15&2.25&2.25&13.96&6.31&6.76&12.61\\
\midrule
CodeNet & success &41.84&7.65&11.73&12.76&2.55&2.55&8.16&0.00&2.55&1.02&4.59&5.61&7.14&2.04\\
& compile &10.71&56.12&55.61&18.88&77.04&79.08&37.76&98.47&78.57&84.69&37.24&63.27&52.04&50.51\\
& runtime &43.37&20.41&19.90&34.18&10.20&8.16&31.63&0.51&13.27&7.65&37.76&14.80&20.92&27.55\\
& incorrect &4.08&15.82&12.76&34.18&10.20&10.20&22.45&1.02&5.61&6.63&20.41&16.33&19.90&19.90\\
\bottomrule
\end{tabular}
}
\end{table*}


\begin{table*}[t]
\centering
\scriptsize
\caption{Detailed execution metrics for translations from Java to Go.}
\resizebox{\textwidth}{!}{%
\begin{tabular}{llrrrrrrrrrrrrrr}
\toprule
Dataset & Metric & Cds MD&Cds RM&DM MD&DP VT&DP RM&DMx MD&L3 MD&Mis MD&Mix RM&Mix MD&P3 VT&P3 RM&P3 MD&Last MD\\
\midrule
All
&success&47.13&6.22&7.89&4.07&0.00&0.24&12.44&1.67&4.07&5.02&1.91&13.16&12.44&0.72\\
&compile&33.01&78.23&74.64&70.81&98.80&97.61&67.22&92.82&86.84&86.84&83.01&73.68&70.33&94.98\\
&runtime&3.59&2.87&3.11&5.50&0.24&0.24&3.83&0.96&2.87&2.39&2.87&1.20&4.07&0.00\\
&incorrect&16.27&12.68&14.35&19.62&0.96&1.91&16.51&4.55&6.22&5.74&12.20&11.96&13.16&4.31\\
\midrule
AVATAR
&success&42.34&4.95&6.76&4.50&0.00&0.45&10.81&0.90&3.15&4.50&0.90&9.91&11.26&0.45\\
&compile&37.39&80.18&75.68&71.62&98.20&96.85&72.97&92.34&88.74&89.64&83.33&74.32&72.52&96.40\\
&runtime&4.95&2.25&3.15&7.21&0.00&0.00&3.60&1.35&3.15&2.70&3.15&0.90&3.15&0.00\\
&incorrect&15.32&12.61&14.41&16.67&1.80&2.70&12.61&5.41&4.95&3.15&12.61&14.86&13.06&3.15\\
\midrule
CodeNet
&success&52.55&7.65&9.18&3.57&0.00&0.00&14.29&2.55&5.10&5.61&3.06&16.84&13.78&1.02\\
&compile&28.06&76.02&73.47&69.90&99.49&98.47&60.71&93.37&84.69&83.67&82.65&72.96&67.86&93.37\\
&runtime&2.04&3.57&3.06&3.57&0.51&0.51&4.08&0.51&2.55&2.04&2.55&1.53&5.10&0.00\\
&incorrect&17.35&12.76&14.29&22.96&0.00&1.02&20.92&3.57&7.65&8.67&11.73&8.67&13.27&5.61\\
\bottomrule
\end{tabular}
}
\end{table*}

\begin{table*}[t]
\centering
\scriptsize
\caption{Detailed execution metrics for translations from Java to Python.}
\resizebox{\textwidth}{!}{%
\begin{tabular}{llrrrrrrrrrrrrrr}
\toprule
Dataset&Metric&14 values\\
\midrule
All&success&60.77&17.94&24.40&16.75&14.35&4.78&32.30&28.23&13.88&16.51&7.89&22.01&30.62&22.97\\
&compile&0.24&28.71&6.94&13.88&21.53&36.36&3.59&3.83&11.24&11.24&10.77&34.45&8.37&3.83\\
&runtime&29.90&29.67&38.28&37.08&38.28&18.66&44.98&54.07&52.39&54.55&47.85&32.78&41.87&58.13\\
&incorrect&9.09&23.68&30.38&32.30&25.84&40.19&19.14&13.88&22.49&17.70&33.49&10.77&19.14&15.07\\
\midrule
AVATAR&success&66.67&13.96&24.32&17.12&8.11&1.80&28.83&26.58&11.26&11.71&7.21&18.02&26.13&15.77\\
&compile&0.00&31.08&6.76&13.51&20.72&44.59&3.60&3.15&14.41&12.16&11.26&40.09&10.36&2.25\\
&runtime&22.52&27.48&34.68&34.68&43.24&20.72&43.69&57.66&52.70&60.81&48.20&30.18&39.64&65.32\\
&incorrect&10.81&27.48&34.23&34.68&27.93&32.88&23.87&12.61&21.62&15.32&33.33&11.71&23.87&16.67\\
\midrule
CodeNet&success&54.08&22.45&24.49&16.33&21.43&8.16&36.22&30.10&16.84&21.94&8.67&26.53&35.71&31.12\\
&compile&0.51&26.02&7.14&14.29&22.45&27.04&3.57&4.59&7.65&10.20&10.20&28.06&6.12&5.61\\
&runtime&38.27&32.14&42.35&39.80&32.65&16.33&46.43&50.00&52.04&47.45&47.45&35.71&44.39&50.00\\
&incorrect&7.14&19.39&26.02&29.59&23.47&48.47&13.78&15.31&23.47&20.41&33.67&9.69&13.78&13.27\\
\bottomrule
\end{tabular}
}
\end{table*}


\begin{table*}[t]
\centering
\scriptsize
\resizebox{\textwidth}{!}{%
\begin{tabular}{llrrrrrrrrrrrrrr}
\toprule
Dataset&Metric&\multicolumn{14}{c}{Models}\\
\midrule
All&success&43.30&8.61&9.33&8.37&0.00&0.00&15.55&5.50&0.24&0.72&0.72&11.00&11.24&0.72\\
&compile&19.86&67.46&65.55&65.79&97.13&97.61&55.50&85.65&96.89&92.58&88.52&72.01&66.99&89.23\\
&runtime&29.67&15.79&17.94&15.07&0.48&0.24&23.21&6.22&0.48&3.35&3.83&12.92&14.83&3.83\\
&incorrect&7.18&8.13&7.18&10.77&2.39&2.15&5.74&2.63&2.39&3.35&6.94&4.07&6.94&6.22\\
\midrule
AVATAR&success&39.64&4.95&5.86&5.86&0.00&0.00&12.61&1.80&0.00&0.90&0.00&8.11&5.86&0.45\\
&compile&22.52&71.17&67.12&72.07&96.85&98.20&58.56&92.79&98.20&96.40&93.69&75.68&72.97&93.69\\
&runtime&28.83&15.32&18.92&13.06&0.45&0.00&22.52&3.60&0.00&1.35&1.80&11.71&13.51&1.35\\
&incorrect&9.01&8.56&8.11&9.01&2.70&1.80&6.31&1.80&1.80&1.35&4.50&4.50&7.66&4.50\\
\midrule
CodeNet&success&47.45&12.76&13.27&11.22&0.00&0.00&18.88&9.69&0.51&0.51&1.53&14.29&17.35&1.02\\
&compile&16.84&63.27&63.78&58.67&97.45&96.94&52.04&77.55&95.41&88.27&82.65&67.86&60.20&84.18\\
&runtime&30.61&16.33&16.84&17.35&0.51&0.51&23.98&9.18&1.02&5.61&6.12&14.29&16.33&6.63\\
&incorrect&5.10&7.65&6.12&12.76&2.04&2.55&5.10&3.57&3.06&5.61&9.69&3.57&6.12&8.16\\
\bottomrule
\end{tabular}}
\caption{Detailed execution metrics for translations from Java to Rust.}
\end{table*}

\begin{table*}[t]
\centering
\scriptsize
\resizebox{\textwidth}{!}{%
\begin{tabular}{llrrrrrrrrrrrrrr}
\toprule
Dataset&Metric&\multicolumn{14}{c}{Models}\\
\midrule
All&success&74.31&43.35&40.83&9.17&9.63&5.96&46.56&38.02&9.86&14.45&1.38&36.47&39.45&20.41\\
&compile&5.96&20.64&24.08&22.48&59.86&68.12&19.72&31.34&54.36&53.90&47.71&20.18&16.97&42.89\\
&runtime&7.57&14.45&12.16&25.92&17.66&15.37&13.99&11.75&12.84&13.30&25.46&9.86&10.78&11.47\\
&incorrect&12.16&21.56&22.94&42.43&12.84&10.55&19.72&18.89&22.94&18.35&25.46&33.49&32.80&25.23\\
\midrule
AVATAR&success&62.78&34.53&30.94&5.83&5.38&4.04&33.18&24.22&4.04&10.31&0.90&27.35&27.35&10.31\\
&compile&7.17&25.11&26.46&28.70&67.71&74.89&26.91&40.36&62.33&60.54&50.22&23.32&21.52&51.57\\
&runtime&13.45&18.39&16.14&28.25&14.80&12.56&17.49&13.00&13.45&12.56&25.56&15.70&16.14&13.45\\
&incorrect&16.59&21.97&26.46&37.22&12.11&8.52&22.42&22.42&20.18&16.59&23.32&33.63&34.98&24.66\\
\midrule
BitHacks&success&78.57&35.71&28.57&0.00&28.57&0.00&35.71&28.57&14.29&7.14&0.00&28.57&28.57&0.00\\
&compile&0.00&35.71&42.86&28.57&57.14&78.57&21.43&42.86&42.86&57.14&35.71&14.29&21.43&42.86\\
&runtime&7.14&0.00&0.00&7.14&0.00&0.00&0.00&0.00&14.29&21.43&21.43&7.14&0.00&14.29\\
&incorrect&14.29&28.57&28.57&64.29&14.29&21.43&42.86&28.57&28.57&14.29&42.86&50.00&50.00&42.86\\
\midrule
CodeNet&success&86.93&53.77&52.76&13.57&13.07&8.54&62.31&54.31&16.08&19.60&2.01&47.24&53.77&33.17\\
&compile&5.03&14.57&20.10&15.08&51.26&59.80&11.56&20.30&46.23&46.23&45.73&17.09&11.56&33.17\\
&runtime&1.01&11.06&8.54&24.62&22.11&19.60&11.06&11.17&12.06&13.57&25.63&3.52&5.53&9.05\\
&incorrect&7.04&20.60&18.59&46.73&13.57&12.06&15.08&14.21&25.63&20.60&26.63&32.16&29.15&24.62\\
\bottomrule
\end{tabular}}
\caption{Detailed execution metrics for translations from Python to C\#.}
\end{table*}


\begin{table*}[t]
\centering
\scriptsize
\resizebox{\textwidth}{!}{%
\begin{tabular}{llrrrrrrrrrrrrrr}
\toprule
Dataset & Metric & \multicolumn{14}{c}{Models}\\
\midrule
All & success &36.24&7.57&5.73&4.59&0.00&0.23&11.01&3.23&7.80&3.21&2.52&11.24&10.55&0.69\\
& compile &35.32&78.67&78.90&70.87&97.71&98.39&76.83&91.01&83.49&90.60&79.59&78.21&79.82&94.95\\
& runtime &14.45&4.59&3.90&3.44&1.15&0.23&3.90&2.76&1.38&1.38&5.96&2.52&4.36&0.92\\
& incorrect &13.99&9.17&11.47&21.10&1.15&1.15&8.26&3.00&7.34&4.82&11.93&8.03&5.28&3.44\\
\midrule
AVATAR & success &29.60&3.59&3.14&2.24&0.00&0.00&8.07&1.79&3.59&1.35&0.45&6.28&5.38&0.45\\
& compile &39.01&80.72&83.86&74.44&97.76&99.10&77.13&92.83&91.03&95.07&83.41&81.17&84.75&96.41\\
& runtime &15.25&4.93&1.79&4.93&0.45&0.00&4.93&1.79&1.35&0.90&4.93&3.14&4.04&0.90\\
& incorrect &16.14&10.76&11.21&18.39&1.79&0.90&9.87&3.59&4.04&2.69&11.21&9.42&5.83&2.24\\
\midrule
BitHacks & success &42.86&7.14&0.00&0.00&0.00&0.00&14.29&7.14&14.29&0.00&0.00&28.57&28.57&0.00\\
& compile &35.71&78.57&92.86&71.43&100.0&100.0&85.71&85.71&78.57&92.86&78.57&71.43&50.00&92.86\\
& runtime &7.14&0.00&0.00&0.00&0.00&0.00&0.00&0.00&0.00&0.00&0.00&0.00&0.00&0.00\\
& incorrect &14.29&14.29&7.14&28.57&0.00&0.00&0.00&7.14&7.14&7.14&21.43&0.00&21.43&7.14\\
\midrule
CodeNet & success &43.22&12.06&9.05&7.54&0.00&0.50&14.07&4.57&12.06&5.53&5.03&15.58&15.08&1.01\\
& compile &31.16&76.38&72.36&66.83&97.49&97.49&75.88&89.34&75.38&85.43&75.38&75.38&76.38&93.47\\
& runtime &14.07&4.52&6.53&2.01&2.01&0.50&3.02&4.06&1.51&2.01&7.54&2.01&5.03&1.01\\
& incorrect &11.56&7.04&12.06&23.62&0.50&1.51&7.04&2.03&11.06&7.04&12.06&7.04&3.52&4.52\\
\bottomrule
\end{tabular}}
\caption{Detailed execution metrics for translations from Python to Go.}
\end{table*}

\begin{table*}[t]
\centering
\scriptsize
\resizebox{\textwidth}{!}{%
\begin{tabular}{llrrrrrrrrrrrrrr}
\toprule
Dataset & Metric & \multicolumn{14}{c}{Models}\\
\midrule
All & success &72.71&44.50&44.95&22.48&15.37&10.32&52.75&36.87&18.81&21.56&11.01&50.23&49.31&22.71\\
& compile &4.13&18.12&19.50&25.00&53.67&58.72&16.74&28.80&54.13&48.85&42.89&21.10&19.72&44.95\\
& runtime &8.26&15.83&16.28&15.60&13.30&14.91&14.91&17.74&11.24&14.45&19.04&11.93&14.45&19.50\\
& incorrect &14.91&21.56&19.27&36.93&17.66&16.06&15.60&16.59&15.83&15.14&27.06&16.74&16.51&12.84\\
\midrule
AVATAR & success &60.99&30.04&32.29&16.14&6.73&5.38&38.57&24.66&8.97&9.42&4.48&36.77&35.43&16.59\\
& compile &5.38&22.87&21.97&27.35&58.74&68.16&16.59&33.18&64.13&54.26&46.19&28.70&23.32&47.09\\
& runtime &10.76&16.59&18.39&19.73&14.35&13.00&20.63&19.73&12.56&18.39&23.32&12.56&18.39&18.39\\
& incorrect &22.87&30.49&27.35&36.77&20.18&13.45&24.22&22.42&14.35&17.94&26.01&21.97&22.87&17.94\\
\midrule
BitHacks & success &50.00&28.57&35.71&14.29&7.14&7.14&35.71&50.00&21.43&14.29&28.57&50.00&50.00&35.71\\
& compile &7.14&28.57&35.71&50.00&71.43&64.29&35.71&35.71&64.29&78.57&42.86&14.29&21.43&64.29\\
& runtime &28.57&35.71&28.57&0.00&14.29&28.57&7.14&7.14&7.14&0.00&0.00&7.14&7.14&0.00\\
& incorrect &14.29&7.14&0.00&35.71&7.14&0.00&21.43&7.14&7.14&7.14&28.57&28.57&21.43&0.00\\
\midrule
CodeNet & success &87.44&61.81&59.80&30.15&25.63&16.08&69.85&49.75&29.65&35.68&17.09&65.33&64.82&28.64\\
& compile &2.51&12.06&15.58&20.60&46.73&47.74&15.58&23.35&42.21&40.70&39.20&13.07&15.58&41.21\\
& runtime &4.02&13.57&13.07&12.06&12.06&16.08&9.05&16.24&10.05&11.06&15.58&11.56&10.55&22.11\\
& incorrect &6.03&12.56&11.56&37.19&15.58&20.10&5.53&10.66&18.09&12.56&28.14&10.05&9.05&8.04\\
\bottomrule
\end{tabular}}
\caption{Detailed execution metrics for translations from Python to Java.}
\end{table*}


\begin{table*}[t]
\centering
\scriptsize
\resizebox{\textwidth}{!}{%
\begin{tabular}{llrrrrrrrrrrrrrr}
\toprule
Dataset & Metric & \multicolumn{14}{c}{Models}\\
\midrule
All & success &61.01&18.58&16.06&10.32&0.00&0.00&29.59&16.82&0.69&1.38&2.52&22.25&22.25&6.42\\
& compile &17.43&66.28&67.20&64.22&98.17&97.25&47.71&65.21&96.10&94.95&88.30&66.28&63.30&86.93\\
& runtime &12.84&8.72&9.63&13.53&0.00&0.69&15.14&7.60&1.38&1.61&2.52&5.05&6.19&2.29\\
& incorrect &8.72&6.42&7.11&11.93&1.83&2.06&7.57&10.37&1.83&2.06&6.65&6.42&8.26&4.36\\
\midrule
AVATAR & success &47.09&11.21&9.42&4.93&0.00&0.00&19.73&4.04&0.00&0.45&0.45&12.11&13.90&0.90\\
& compile &21.08&72.65&75.34&70.85&97.76&98.65&50.67&78.03&96.86&96.86&92.83&74.44&68.16&94.62\\
& runtime &17.94&9.42&9.87&13.90&0.00&0.00&19.73&6.28&1.35&0.90&1.35&5.38&8.52&1.79\\
& incorrect &13.90&6.73&5.38&10.31&2.24&1.35&9.87&11.66&1.79&1.79&5.38&8.07&9.42&2.69\\
\midrule
BitHacks & success &50.00&14.29&42.86&7.14&0.00&0.00&28.57&14.29&7.14&7.14&7.14&28.57&14.29&0.00\\
& compile &28.57&78.57&50.00&57.14&100.0&100.0&50.00&64.29&92.86&85.71&85.71&64.29&64.29&92.86\\
& runtime &21.43&0.00&0.00&28.57&0.00&0.00&21.43&14.29&0.00&0.00&0.00&0.00&7.14&7.14\\
& incorrect &0.00&7.14&7.14&7.14&0.00&0.00&0.00&7.14&0.00&7.14&7.14&7.14&14.29&0.00\\
\midrule
CodeNet & success &77.39&27.14&21.61&16.58&0.00&0.00&40.70&31.47&1.01&2.01&4.52&33.17&32.16&13.07\\
& compile &12.56&58.29&59.30&57.29&98.49&95.48&44.22&50.76&95.48&93.47&83.42&57.29&57.79&77.89\\
& runtime &6.53&8.54&10.05&12.06&0.00&1.51&9.55&8.63&1.51&2.51&4.02&5.03&3.52&2.51\\
& incorrect &3.52&6.03&9.05&14.07&1.51&3.02&5.53&9.14&2.01&2.01&8.04&4.52&6.53&6.53\\
\bottomrule
\end{tabular}}
\caption{Detailed execution metrics for translations from Python to Rust.}
\label{tab:python-rust}
\end{table*}

\newpage

\end{document}